\documentclass[12pt,preprint]{aastex}
\usepackage{amssymb}
\usepackage{graphicx}
\usepackage{amsmath}
\usepackage{natbib}
\usepackage{epsfig}
\usepackage{epstopdf}
\newcommand{\asec} {\mbox{$^{\prime \prime}$} }

\usepackage{mathrsfs}
\usepackage{amsfonts}

\slugcomment{Version Mar 20th, 2010. To be submitted to ApJ}
\shorttitle{HDP Quasars in R06 and E94} \shortauthors{Hao et al.}

\begin{document}

\title{Hot-Dust-Poor Quasars in Mid-Infrared and Optically Selected Samples}

\author{Heng Hao\altaffilmark{1}, Martin Elvis\altaffilmark{1},
Francesca Civano\altaffilmark{1} \& Andy Lawrence\altaffilmark{2,
3}}

\altaffiltext{1}{Harvard-Smithsonian Center for Astrophysics, 60
Garden Street, Cambridge, MA 02138}

\altaffiltext{2}{Institute for Astronomy, University of Edinburgh,
Royal Observatory, Blackford Hill, Edinburgh, EH9 3HJ, UK}

\altaffiltext{3}{Visiting Scientist, Kavli Institute for Particle
Astrophysics and Cosmology (KIPAC), Stanford University, Stanford,
CA 94309, USA}

\email{hhao@cfa.harvard.edu, elvis@cfa.harvard.edu}

\begin{abstract}
We show that the Hot-Dust-Poor (HDP) quasars, originally found in
the X-ray selected XMM-COSMOS type 1 AGN sample, are just as common
in two samples selected at optical/infrared wavelengths: the
Richards et al. Spitzer/SDSS sample ($8.7\%\pm2.2\%$), and the
PG-quasar dominated sample of Elvis et al. ($9.5\%\pm 5.0\%$). The
properties of the HDP quasars in these two samples are consistent
with the XMM-COSMOS sample, except that, at the $99\%~(\sim
2.5\sigma)$ significance, a larger proportion of the HDP quasars in
the Spitzer/SDSS sample have weak host galaxy contributions,
probably due to the selection criteria used. Either the host-dust is
destroyed (dynamically or by radiation), or is offset from the
central black hole due to recoiling. Alternatively, the universality
of HDP quasars in samples with different selection methods and the
continuous distribution of dust covering factor in type 1 AGNs,
suggest that the range of SEDs could be related to the range of
tilts in warped fueling disks, as in the model of Lawrence and Elvis
(2010), with HDP quasars having relatively small warps.
\end{abstract}


\section{Introduction}
Emission from hot ($\sim$ 1500K) dust is so characteristic of AGNs
and quasars (e.g. Suganuma et al. 2006) that the strong 1--3$\mu m$
emission from this dust has often been used to select AGN samples
(e.g. Miley et al. 1985, Lacy et al. 2004, 2007, Stern et al. 2005,
Donley et al. 2008).

However, we recently reported (Hao et al. 2010) that, for 6\% (at
$z<2$) to 20\% (at $2<z<3.5$) of the quasars in the XMM-COSMOS type
1 AGN sample (Brusa et al. 2010; Elvis et al. 2011), their $1-3\mu
m$ emission are two to four times weaker than typical type 1 AGN
(e.g., Elvis et al. 1994), although they have normal optical
`big-blue-bump' slopes indicative of standard accretion disk
emission. We dubbed these `hot-dust-poor' (HDP) quasars. Their
implied dust `torus' covering factor is $\sim$2\% to 30\%, well
below the 75\% predicted by the unified model (e.g. Krolik \&
Begelman 1988). These sources lie at one extreme of the distribution
of AGN SEDs found in the XMM-COSMOS sample. Their lack of hot dust
emission uncovers a continuum that appears to be the continuation of
the accretion disk, which implies a disk size of $\sim 10^4$
Schwarzschild radii, an order of magnitude beyond the gravitational
instability radius.

\begin{figure}
\includegraphics[angle=0,width=0.48\textwidth]{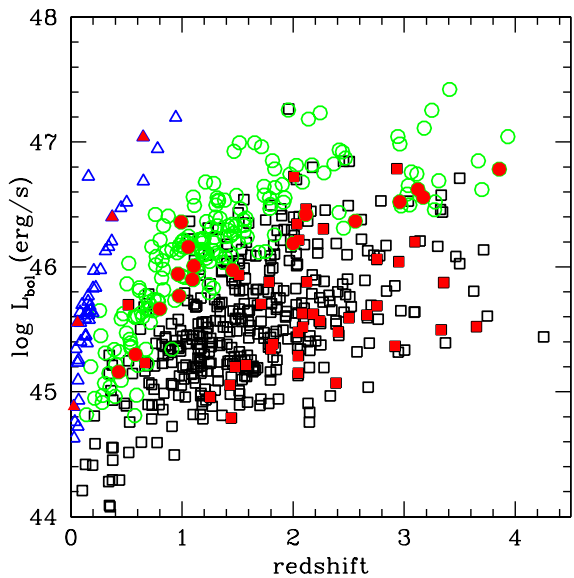}
\caption{The bolometric luminosity and redshift of the XMM-COSMOS
(squares), R06 (circles) and E94 (triangles) sample. The red solid
points are the HDP quasars from these samples. The E94 quasars show
a narrow distribution, as they are chosen to be the very brightest
quasars in the sky. $L_{bol}$ is calculated as in Hao et al. (2010)
by integrating from rest frame 24 $\mu m$ to the Lyman limit
(912\AA) and using the WMAP 5-year cosmology: $H_0=71
~~km~s^{-1}Mpc^{-1}$, $\Omega_M=0.26$ and $\Omega_\Lambda=0.74$
(Komatsu et al., 2009). \label{zLbol}}
\end{figure}

The study of Hao et al. (2010) used an X-ray selected sample. To
check whether weak dust emission is a property of quasars in
general, as opposed to X-ray selected quasars, in this paper we
examine a large optical/infrared (IR) selected sample, that of
Richards et al. (2006; hereafter R06). As shown in
Figure~\ref{zLbol}, R06 also covers a different part of the
luminosity/redshift (L,z) plane from the COSMOS-XMM sample, which
will help us to test the preliminary claim of Hao et al. (2010) that
the HDP fraction changes with redshift but not with luminosity. We
also examine the Elvis et al. (1994) sample of low-redshift quasars
(hereafter E94). Figure~\ref{zLbol} shows that the three samples
together have a good coverage of the (L,z) plane.

\section{Type 1 AGN samples}
Hao et al. (2010) selected HDP quasars by their position in a plot
of optical (OPT, 0.3--1$\mu m$) versus near-infrared (NIR, 1--3$\mu
m$) slopes. In this paper, we use the samples of R06 and E94
supplemented by new near-IR UKIDSS data, which gives good enough
photometry to define these slopes.

\subsection{R06 sample}

The R06 sample consists of 259 {\it Spitzer} sources identified with
SDSS quasars in four discrete degree-scale fields, and is,
therefore, mid-IR and optically selected. We also considered the
similar Hatziminaoglou et al. (2008) sample of 278 quasars and found
that all but 65 are also in R06 and only 39 of these 65 have Spitzer
photometry. We therefore only use R06. The redshift range covered is
$z=0.14 - 5.2$ with 93\% being at $z<3$ (Figure~\ref{zLbol}).

Most of the data in R06 paper is from SDSS, Spitzer, and 2MASS
photometry. These data were not corrected for host galaxy
contributions. As R06 is an IR and optically selected sample, the
host galaxy contribution should typically be small. As the galaxy
SED in the IR and optical is completely different from the typical
quasar SED, and IR and optical color selection picks out sources
with spectral indices similar to those of a typical quasar, sources
with more galaxy contribution are selected against (e.g. Gregg et
al. 2002, Richards et al. 2003). Most (215/259) of the R06 sources
did not have 2MASS J H K photometry. These bands are essential to
make reliable NIR slope estimates for quasars at $z\lesssim0.65$,
and to make reliable OPT slope estimates for quasars at
$z\gtrsim1.3$. We therefore cross-matched the R06 sample with the
UKIDSS database using the online WFCAM science archive. We found
additional data for 98 quasars at K and for 45 quasars at J,
producing a final subset of 195 R06 quasars that we can use for this
study.

The UKIDSS data come from the Deep Extragalactic Survey (DXS) fields
covering the Lockman Hole and ELAIS-N1. The UKIDSS survey is defined
in Lawrence et al (2007). It uses the UKIRT Wide Field Camera
(WFCAM; Casali et al, 2007). The photometric system is described in
Hewett et al (2006), and the calibration is described in Hodgkin et
al. (2009). The pipeline processing  and science archive are
described in Irwin et al (2010, in prep) and Hambly et al (2008).
The magnitudes we use here are ``aper3'' magnitudes, which come from
a measurement through a 2 \asec software aperture, corrected to a
total magnitude using the point spread function measured in the
field. We have not corrected these magnitudes for host galaxy
contribution, but we could see that nearly all of these objects only
have small extended source contribution, because the UKIDSS database
provides a wide range of aperture magnitudes. The magnitudes we use
come from stacked observations spreading over several years. From
individual epochs, we can see that most of these quasars vary only
by a small amount (0.05 mag) within the UKIDSS data, such that it
does not seriously affect our analysis.

\subsection{E94 sample}
The E94 quasar sample consists of both optically-selected quasars
from the bright quasar survey (BQS, PG, Schmidt \& Green, 1983) and
radio-selected (mainly 3C and PKS) quasars. These quasars were
selected to have good signal-to-noise ratio {\it Einstein} X-ray
observations and obtainable IUE UV spectra, hence E94 is biased
towards X-ray bright and blue quasars. E94 contains 42 quasars in
the redshift range $z= 0.025 - 0.94$, with 80\% being at $z<0.3$.
The optical photometry were obtained at the FLWO (F. L. Whipple
Observatory) 24 inch telescope within one week of the MMT FOGS
(Faint Object Grism Spectrograph) observations. The NIR data were
obtained with MMT and IRTF. More details on the observation can be
found in Elvis et al. (1994). As E94 is mainly a local quasar
sample, the E94 SEDs have been corrected for host galaxy
contamination by subtracting the host galaxy template SED based on
the Sbc galaxy model of Coleman, Wu \& Weedman (1980). The
normalization of the galaxy template was calculated by direct
measurements of the host galaxy luminosities.

\section{Mixing Diagram}
To select HDP quasars, we use the mixing diagram technique of Hao et
al. (2010), where we select objects that lie outside the region
formed by mixing a typical quasar SED with a host galaxy SED,
combined with reddening. We use the slope on either side of the
1$\mu$m inflection point, which is where the black-body emission of
the hottest dust rises above that of the accretion disk. We build
the mixing diagram by plotting two rest frame slopes in $log~\nu
L_{\nu}$ versus $log~\nu$ space: $\alpha_{NIR}$ ($3\mu m - 1\mu m$)
versus $\alpha_{OPT}$ ($1\mu m - 3000$\AA). These slopes are derived
from linear fits to the SEDs in these wavelength ranges. The errors
on the slopes are the 1$\sigma$ standard error. We required that at
least three photometry points are used to fit the slopes. 88\% of
R06 quasars have more than the minimum of three photometry points in
the optical and 71\% have more than three points in the
near-infrared. The measurement error on the photometry is used in
the fitting. For both R06 and E94, all these slopes are measured in
the same rest frame wavelength range as the XMM-COSMOS sample.

The mixing diagram is equivalent to a color-color plot, but utilizes
more photometric points and adapts well to a wide range of
redshifts. The mixing diagram is more robust than the normal
color-color plot, because it is less affected by one bad photometry
point and is built in the rest frame without any need for
assumptions about the intrinsic SED shape (k-correction).

\subsection{Robustness of Linear Fitting}
\label{s:robust}

We performed two tests to check if the requirement to have three or
more photometry band is adequate to ensure robust slope
measurements.

\begin{figure*}
\includegraphics[angle=0,width=0.4\textwidth]{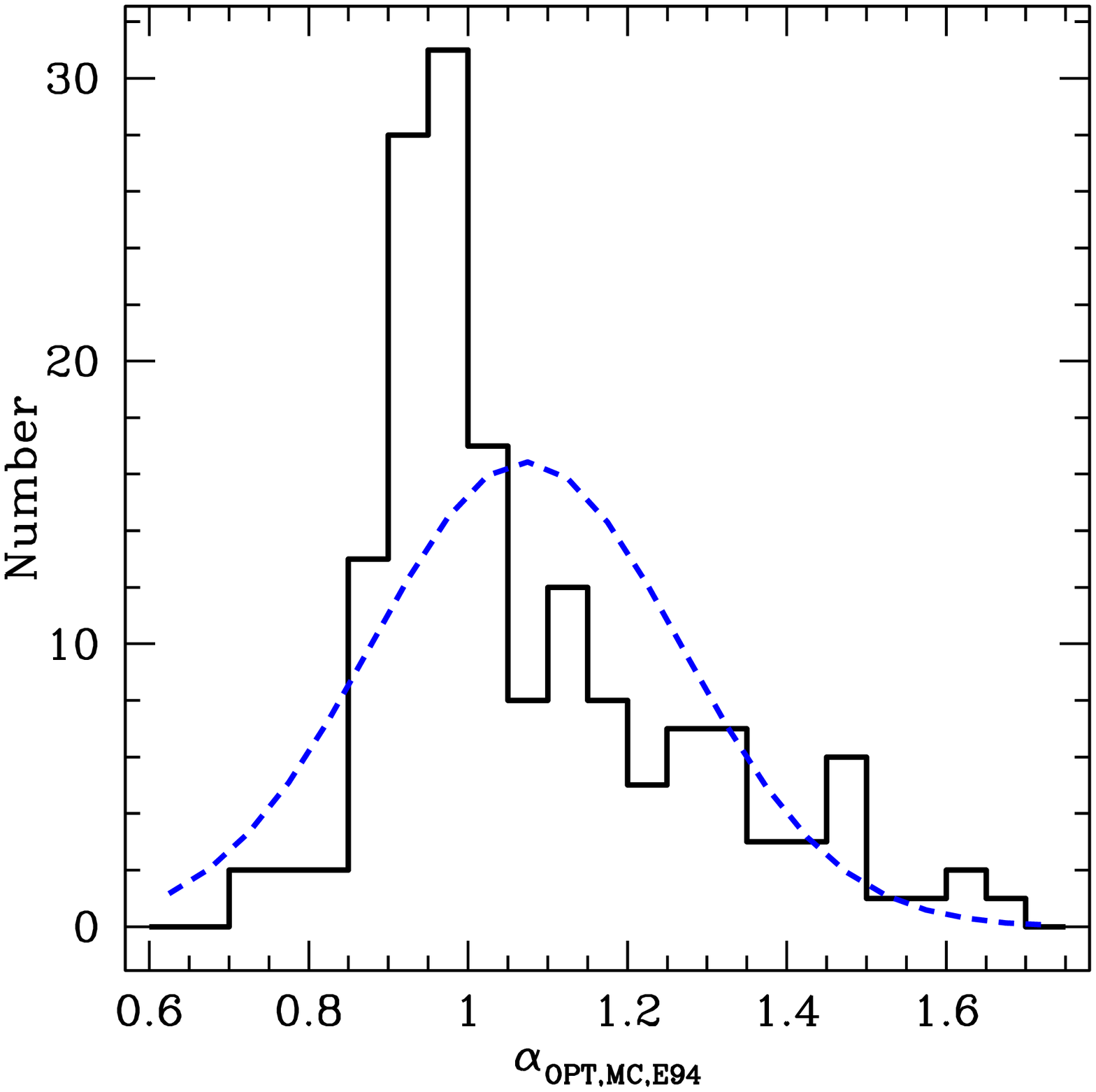}
\includegraphics[angle=0,width=0.4\textwidth]{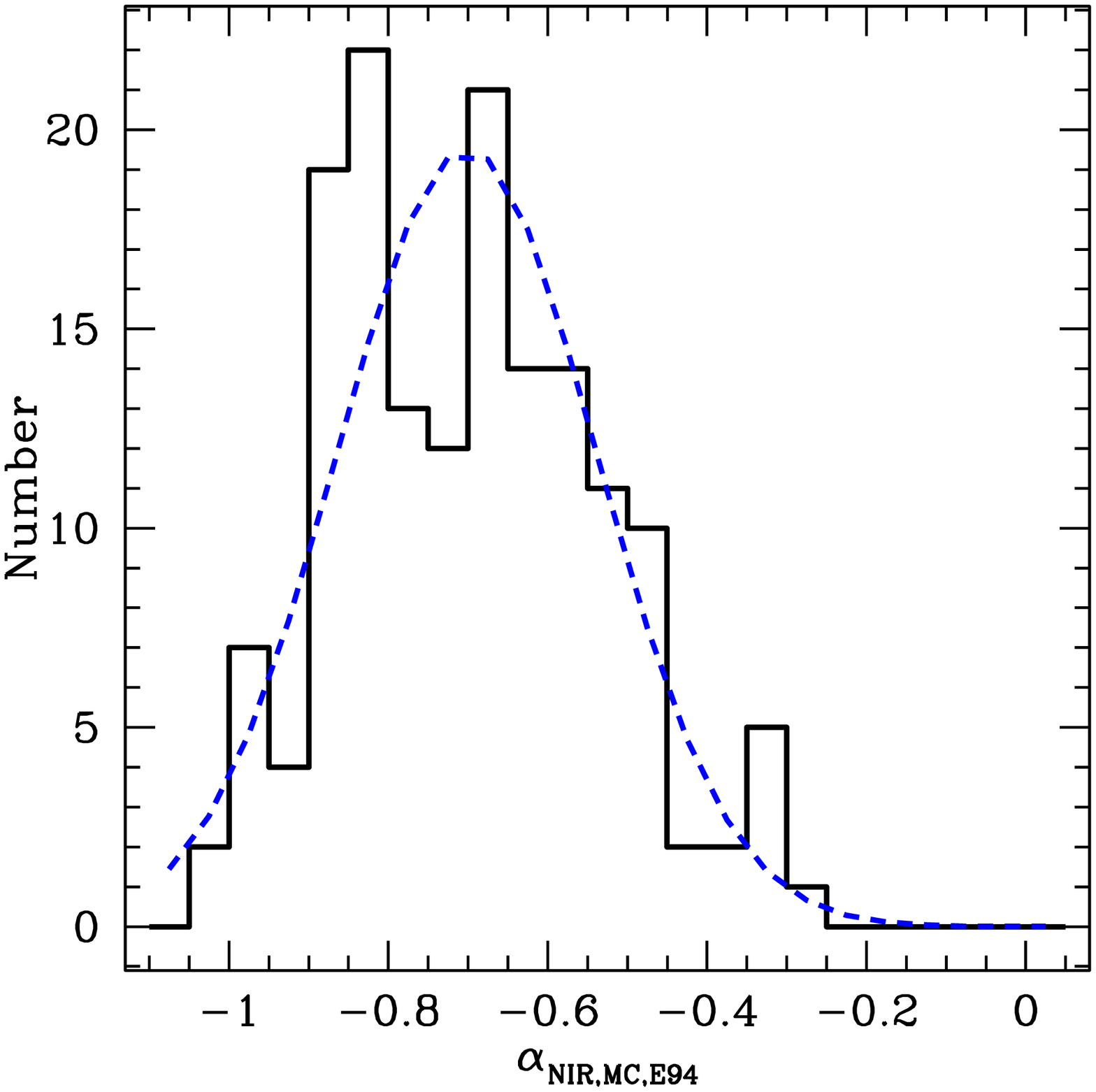}
\caption{Histogram of the E94 slopes with the SDSS-Spitzer sampling
for the R06 quasars: {\em Left:} $\alpha_{OPT}$; {\em Right:}
$\alpha_{NIR}$. The blue dashed lines are the Gaussian fit of the
histogram. \label{slphist}}
\end{figure*}

\begin{figure*}
\includegraphics[angle=0,width=0.48\textwidth]{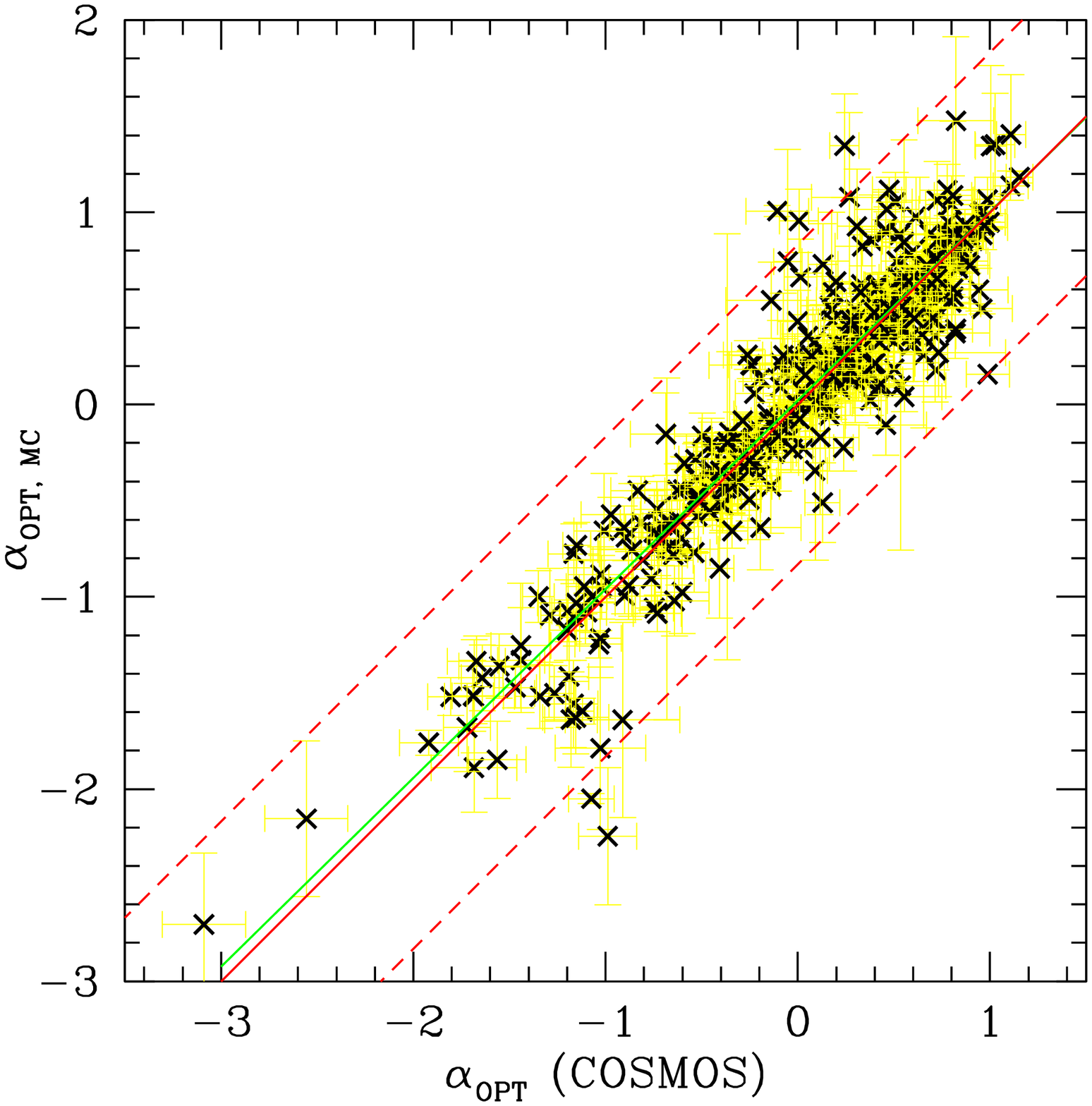}
\includegraphics[angle=0,width=0.48\textwidth]{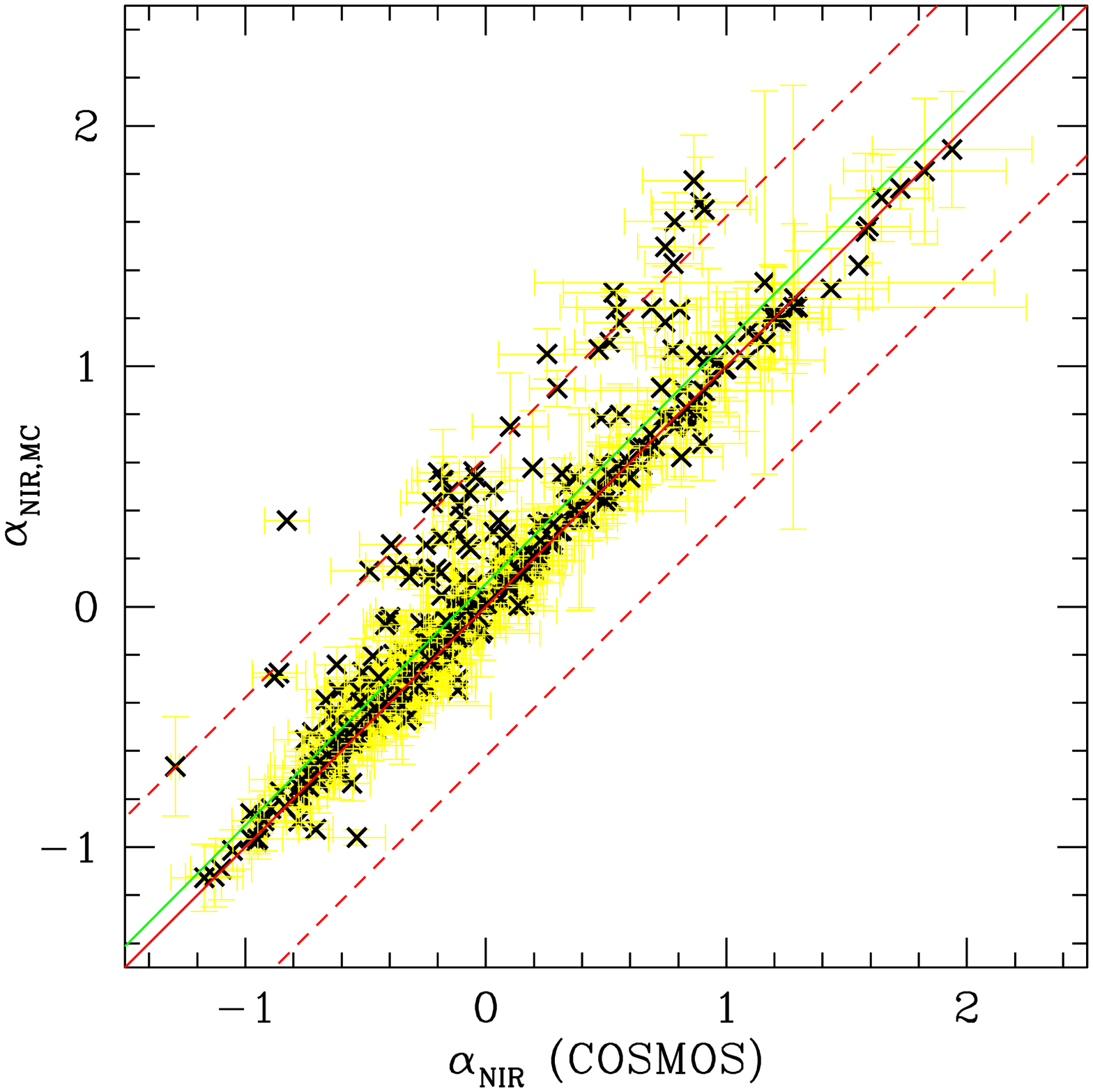}
\caption{Comparison of the simulated SDSS photometry slopes with the
XMM-COSMOS photometry slopes for the XMM-COSMOS quasars: {\em Left:}
$\alpha_{OPT}$; {\em Right:} $\alpha_{NIR}$. The red solid line is
the equal slope line. The green line is the linear fit of the black
points. The two red dashed lines are at the $3\sigma_d$ distance to
the equal slope line. The error bars of the slope measurement are
shown as yellow.\label{slpcp}}
\end{figure*}

\begin{description}
  \item [(1)]~~We use the E94 mean SED as a template. This SED has $\alpha_{OPT}=0.95$
  and $\alpha_{NIR}=-0.72$. We made a Monte Carlo sampling of the E94 SED, with
  the same sampling and error bars as the observed SEDs for each R06 quasar.
  We then measured the simulated slopes, as we did for the observed ones. We
  show the histogram of the simulated slopes in Figure~\ref{slphist}.

  For the simulated optical slopes, the mean is
  $\langle\alpha_{OPT,MC,E94}\rangle=1.07$ ($\sigma_{OPT,MC,E94}=0.19$).
  The E94 optical slope ($\alpha_{OPT}=0.95$) is within 1$\sigma_{OPT,MC,E94}$
  of the mean value. However, when considering the errors ($\sigma$) on each
  Monte Carlo optical slope, we find that only 75\% of the simulated
  quasars lie within 3$\sigma$ of the E94 optical slope, implying significant
  outliers. In fact, a tail toward steep slopes is evident in
  Figure~\ref{slphist} (left). This tail is cause by the curvature
  in the E94 SED in the 0.3-1$\mu$m range. The steepest part of E94
  reaches a slope of $\sim1.5$ between 0.3--0.45$\mu$m ($14.8\lesssim
  log\nu \lesssim 15$). If the simulated photometry points happen to lie
  only in this region, then the simulated slope will have a larger value
  than the $\alpha_{OPT}=0.95$ calculated over the full 0.3-1$\mu$m range.
  This is a limitation of using just the E94 template, which we will
  address next.

  For the corresponding simulated infrared slopes, the mean is
  $\langle\alpha_{NIR,MC,E94}\rangle=-0.70$ ($\sigma_{NIR,MC,E94}=0.17$).
  The E94 infrared slope ($\alpha_{NIR}=-0.72$) is within
  1$\sigma_{NIR,MC,E94}$ of the mean value. When considering the errors
  ($\sigma$) on each Monte Carlo infrared slope, we find that 91\%
  of the simulated quasars lie within 3$\sigma$ of the E94 infrared slope.
  The histogram of the simulated infrared slopes shows two peaks.
  One peak is located at $\sim -0.72$, which is the E94 infrared slope.
  The other peak is located at $\sim -0.85$. As for the tail in
  $\alpha_{OPT,MC,E94}$, this second peak is due to curvature in the E94
  SED in the 1-3$\mu$m range. The steepest part of E94 reaches a slope of
  $\sim -0.9$ at 1.2--2.4$\mu$m ($14.1\lesssim log\nu \lesssim 14.4$).
  If the simulated photometry points happen to lie only in this region,
  the simulated slope will have a large value compared to the E94
  infrared slope calculated over the full 1-3$\mu$m range.

\end{description}
  While the mean of the simulated slopes are same as the E94
  values, the offsets are a cause of concern as they may falsely
  indicate an HDP object. Therefore, we performed a second Monte
  Carlo test using the measured slopes of the COSMOS sources with
  good photometry coverage. This method avoids
  the systematic errors from using a single template SED.

\begin{figure}
\includegraphics[angle=0,width=0.48\textwidth]{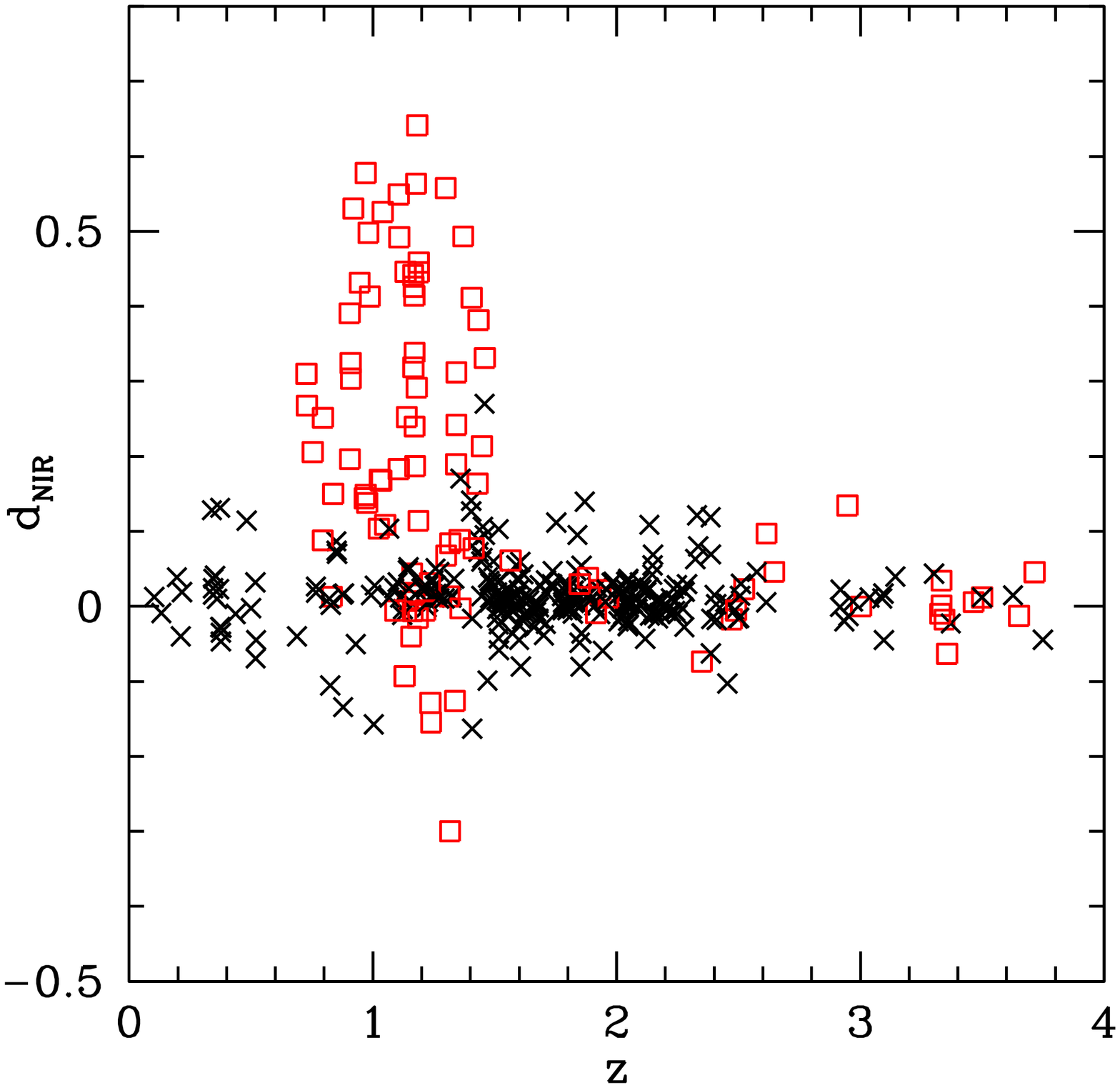}
\caption{The distance of the near-infrared slopes
(Figure~\ref{slpcp}, right) to the equal slope line versus redshift.
The red squares represent the Monte Carlo slope measured with 3 points. 
The black cross represent the Monte Carlo slope measured with more than 3 points
\label{dz}}
\end{figure}

\begin{description}
  \item[(2)]~~To test if the slope measurement is robust for the range of SED shapes
  seen in the XMM-COSMOS type 1 AGN sample, we used the XMM-COSMOS SEDs
  as the template SEDs. The photometry coverage for these SEDs is rich
  in the NIR to OPT range, giving robust slope measurements
  (Elvis et al. 2011, Hao et al. 2010).

  We selected 398 (97\%) XMM-COSMOS type 1 AGNs, that lie within
  $\Delta z=0.1$ of a corresponding quasar in R06 sample. We used
  these 398 quasars to perform a Monte Carlo sampling, matching
  the photometry and error bars to those observed in the corresponding
  R06 quasar. We then measured the resulting $\alpha_{OPT,MC}$
  and $\alpha_{NIR,MC}$, and compared them with the original
  measurements from COSMOS. The results are shown in Figure~\ref{slpcp}.
  A linear fit to $\alpha_{OPT,MC}$ versus $\alpha_{OPT}$
  has a slope of 0.98$\pm$0.0004. The linear fit of $\alpha_{NIR,MC}$
  versus $\alpha_{NIR}$ has a slope of 1.005$\pm$0.0003. Both slopes
  are very close to 1. However there are outliers that emulate HDP
  quasars.

  To investigate these outliers, we defined the distance of each
  data point to the equal slope line as $d_w=(\alpha_{w,MC}-\alpha_w)/\sqrt{2}$
  where $w=OPT, NIR$. A positive value means that the point lies
  above the equal slope line. For the simulated optical slopes,
  the mean of $d_{OPT}$ is 0.015 ($\sigma_{d,OPT}=0.20$), consistent
  with 0 in $1\sigma_{d, OPT}$. For the simulated near-infrared slopes,
  the mean of $d_{NIR}$ is 0.067 ($\sigma_{d,OPT}=0.15$), consistent with 0
  in $1\sigma_{d, NIR}$.

  We use these $\sigma_{d,w}$ to isolate 16 sources that lie
  outside the $3\sigma_{d, NIR}$ lines (red dashed lines), 4\%
  of the total sample. These sources have 3 points in the NIR
  slope estimation and at redshift $0.92\lesssim z\lesssim1.3$
  (see example in Figure~\ref{egSED}, right).
  We plotted the $d_{NIR}$ versus redshift, shown in
  Figure~\ref{dz}. For an E94-like SED shape, the three IRAC bands
  in this redshift range would lie at $>2\mu$m, in the
  upper part of the infrared bump. The lack of a K band measurement
  would then lead to a larger slope estimate. For galaxy-like SED shape
  in this redshift range, lack of K band would lead to a much smaller
  slope estimate. For a mixing of E94 and galaxy SED, which is quite
  flat, the lacking of K band will not affect the slope measurement.
  The total number of simulated sources with 3 points in
  the NIR slope estimation is 95. Thus for the majority (83\%) of the sources
  with 3 photometry points in the NIR slope estimation, the slope estimates
  are still robust.

  In all, for the 96\% simulated sources with 3 or more
  photometry points in NIR, the slope estimation is robust. The NIR
  slope estimate is not quite reliable for sources with only 3 photometry points
  and at redshift $0.92\lesssim z\lesssim1.3$.

\end{description}

In general, for majority of quasars in R06 sample, the slope
measurement we applied is robust for E94 like SED shape or a variety
of SED shapes as in XMM-COSMOS sample, even for quasars with limited
photometry coverage as in R06. Figure~\ref{dz} also shows that the
systematic error on $\alpha_{NIR}$ is 0.07 for sources using 4
or more photometry points, which is small compared to the measurment
errors. For sources using only 3 photometry points in the redshift
range 0.92--1.3, the systematic error on $\alpha_{NIR}$  is 0.3, which is comparable in
some cases to the measurement errors. We added this errors to the
measurement error when plotting the mixing diagram of R06.

\subsection{Corrections to the SED}
We need to consider the following several factors, which might
affect the measurement of the slopes.

\subsubsection{Emission Lines}
The major emission lines in the rest frame wavelength range
$0.3-3\mu m$ are the Balmer series, the Paschen series, and the Fe
II complex. As the photometry points in the SED are all broad-band,
even for the strongest line, H$\alpha$ (mean equivalent width
257\AA\ for SDSS DR6 quasars, Elvis et al. 2011), the upward bias of
the observed photometry from the continuum reaches only 0.1 dex in
the z band (filter FWHM=1000\AA). For weaker lines, such as H$\beta$
and Fe II, the bias would be $\sim$0.02 dex, which is negligible for
the slope measurements reported here. For consistency, we correct
the R06 photometry for the emission lines as we did for the
XMM-COSMOS sample (Elvis et al. 2011).

For the E94 sample, the emission line contribution has already been
corrected for in the reported SEDs (details in Elvis et al. 1994).

\subsubsection{Variability}

Most AGNs vary in their optical continuum flux by $\sim$10\% on
timescales of months to years (Vanden Berk et al. 2004). The
photometry used to build the SEDs in the R06 sample were taken over
different time periods, which may affect the estimation of slopes.

For the R06 sample, the IRAC detections were made around 2004 (Lacy
et al. 2005, Surace et al. 2005). The SDSS-DR3 photometry including
the optical SDSS bands were taken from 1998 to 2003 and the NIR
2MASS bands were taken from 1997 to 2001 (Schneider et al. 2005).
The UKIDSS NIR photometry has been taken since 2005. We should be
careful in using the UKIDSS data and the rest of the data to measure
the slopes.

For the $\alpha_{NIR}$ estimation, we need to consider sources in
the following redshift ranges respectively:
\begin{enumerate}
\item $z>1.15$: only the IRAC bands lie in the 1--3$\mu$m range. Thus the
$\alpha_{NIR}$ estimation is robust, for the 109 out of the 195 R06
quasars;
\item $0.65<z<1.15$: K and IRAC bands lie in the 1--3$\mu$m range.
16 sources have UKIDSS K-band detection. In no case does the K-band
point appear significantly discontinuous with the IRAC bands.
\item $0.24<z<0.65$: H, K and IRAC bands lie in the 1--3$\mu$m range.
17 sources have 2MASS H band detections. 8 sources have UKIDSS
K-band detection. In no case does the H-band and K-band point appear
significant discontinuous with the IRAC bands.
\item $z<0.24$: All of the J, H, K and IRAC1 bands lie in
the 1--3$\mu$m range. 2 of the 3 sources at $z<0.24$ have UKIDSS J
and K band detections. The SEDs of these 2 sources show no
discontinuity in the SED to cause errors in the slope measurement.
\end{enumerate}

We thus consider the estimates of $\alpha_{NIR}$ to be robust.

For $\alpha_{OPT}$ estimation, similarly, the 3 low redshift
($z<0.24$) sources have only the SDSS data in the 0.3--1$\mu$m
range. Thus for these quasars the $\alpha_{OPT}$ estimation is
robust. For sources at $0.24<z<1.15$, only the J band and SDSS data
lie in the range. 12 sources have UKIDSS J band detection, for which
the effect of J band on the slope measurement is small. For the 109
high redshift ($z>1.15$) sources, the situation is more complicated,
because the UKIDSS J K band and the SDSS or 2MASS H band could be in
the 0.3--1$\mu$m range. However, the effect of variability on the
$\alpha_{OPT}$ does not affect the selection of the HDP quasars,
since the most significant difference of the HDP quasars from the
normal quasars are the $\alpha_{NIR}$.

For the E94 sample, the optical and the ground-based IR data were
generally obtained within about 1 month (Elvis et al. 1994), so that
the variability does not affect the slope measurement in the optical
and IR range used in the mixing diagram.

\begin{figure*}
\includegraphics[angle=0,width=0.48\textwidth]{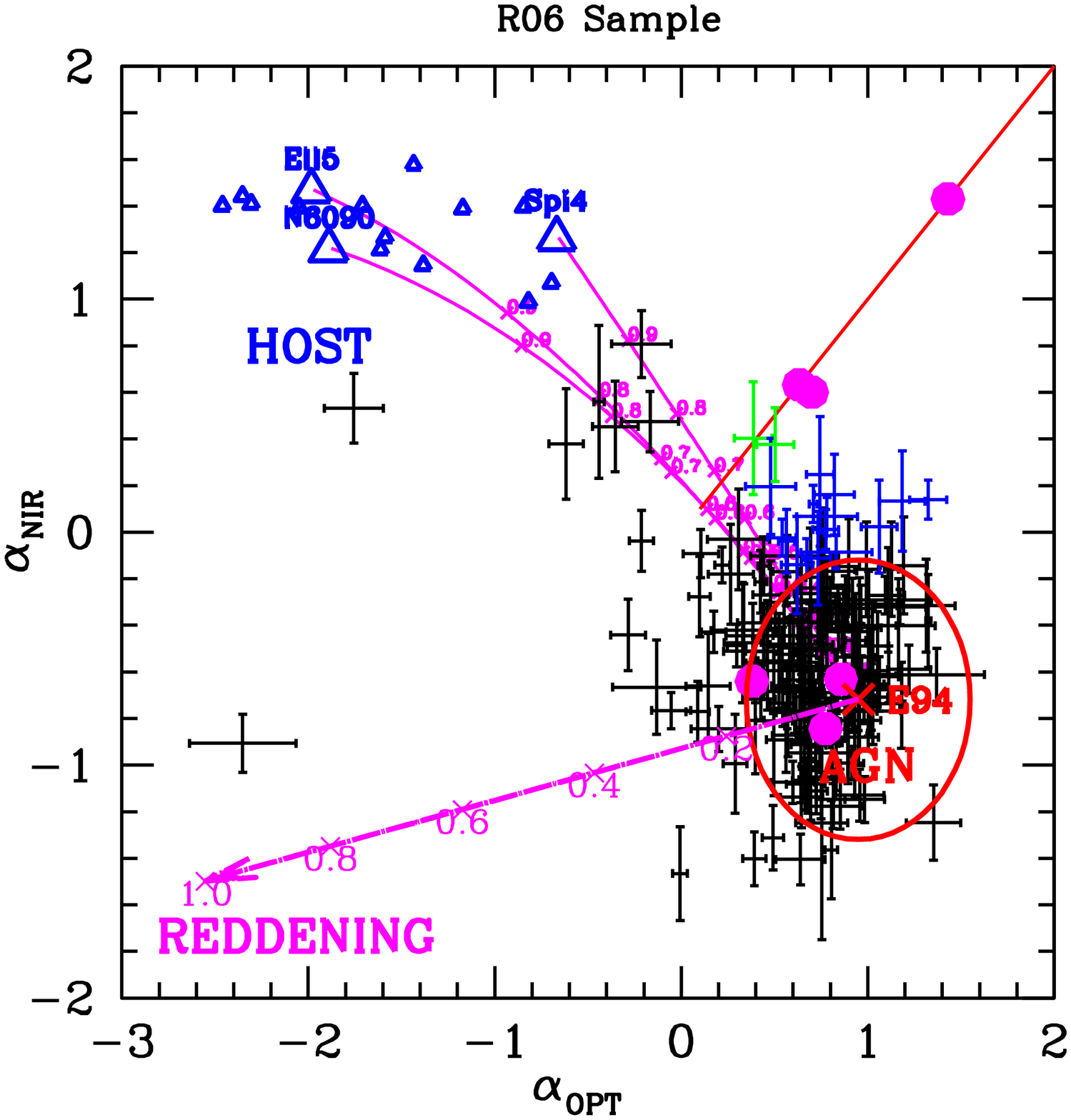}
\includegraphics[angle=0,width=0.48\textwidth]{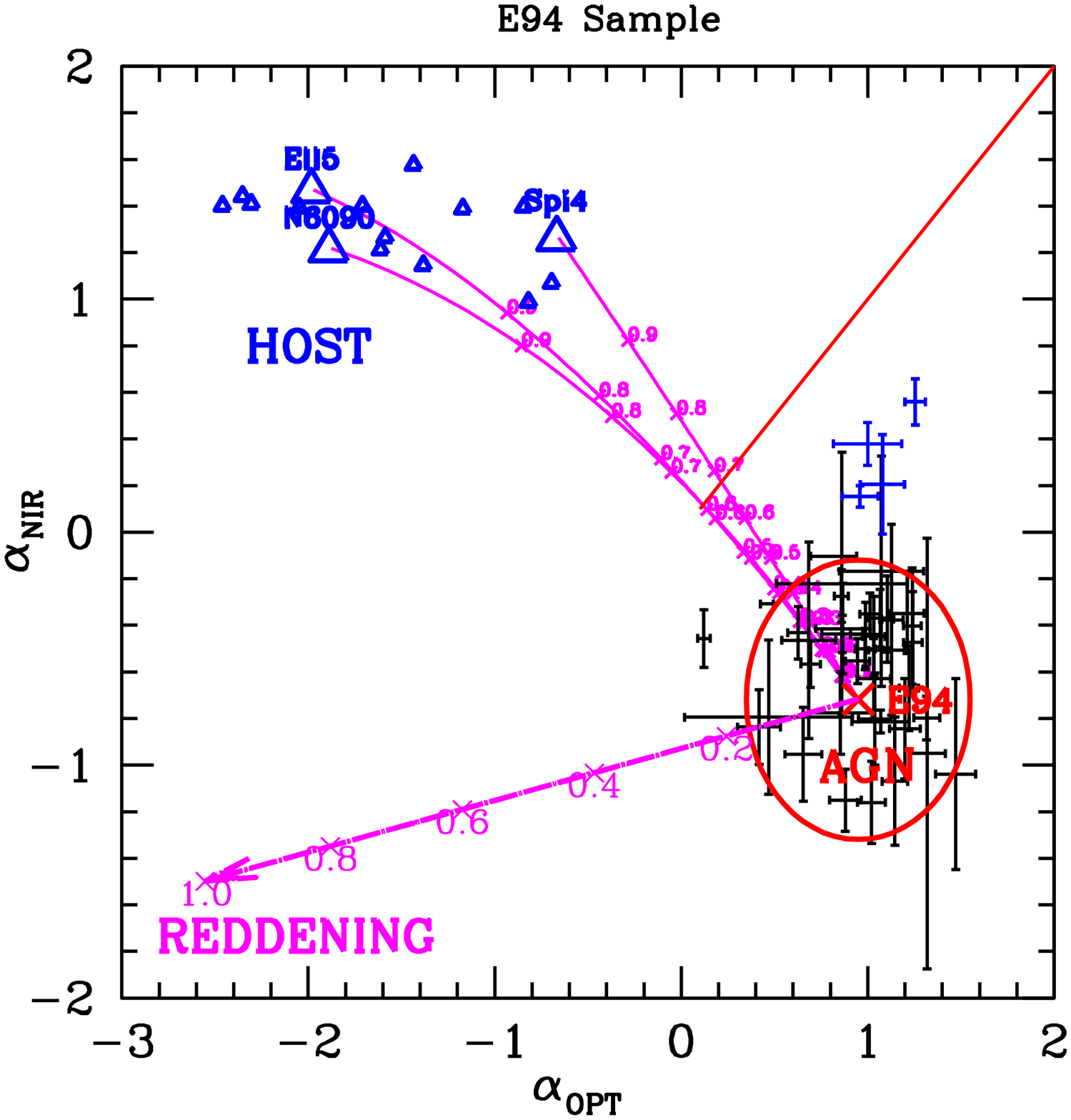}
\caption{Mixing Diagram: {\em Left:} R06 sample; {\em Right:} E94
sample. Red cross and red circle show the E94 mean SED and the
$1\sigma$ dispersion of the E94 sources. The blue triangles show 16
different SWIRE galaxy templates (Polletta et al. 2007). The purple
lines connecting the E94 and the galaxy templates are mixing curves
showing the slopes of different fraction of galaxy and AGN. The
straight purple arrow shows the reddening vector of E94. The
straight red solid line shows the $\alpha_{OPT}=\alpha_{NIR}$ line.
Different colors of the points show different class of the HDP
sources (I--blue, II--green, see text for details). The black
symbols show all the other type 1 AGN in the samples. For R06, the
error bars are adjusted for possible systematic errors due to limit
photometry. The 6 magenta dots in the mixing diagram of R06 sample
show the $z\sim6$ quasars in Jiang et al. (2010) Figure 1.
\label{md}}
\end{figure*}

\subsection{R06 and E94 Mixing Diagram}
Figure \ref{md} shows the Hao et al. (2010) ``mixing diagram'' for
the R06 and E94 samples. For the sources that have galaxy fraction
larger than 0.6 in the mixing diagram, we check the inferred host
galaxy luminosity. Only one quasar in R06 sample has $M<-23$ (-23.8)
at $1\mu$m. Thus the inferred host luminosity is reasonable. There
is clearly a continuous distribution of objects in $\alpha_{NIR}$,
so to select quasars in the HDP region, we need to examine the
dispersion in ``normal'' objects.

We consider all R06 quasars with $0.2<\alpha_{OPT}<1.6$ in order to
exclude the few galaxy- or reddening-dominated sources. They have
mean slopes $\bar{\alpha}_{OPT}=0.76$ (standard deviation
$\sigma_{OPT}=0.25$), and $\bar{\alpha}_{NIR}=-0.53$ (standard
deviation $\sigma_{NIR}=0.37$). The mean of the measurement error of
$\alpha_{OPT}$ is $Err_{OPT}=0.10$, which indicates the intrinsic
dispersion is $\sigma_{INT,
OPT}=\sqrt{\sigma_{OPT}^2-Err_{OPT}^2}=0.23$. The mean of the
measurement error of $\alpha_{NIR}$ is $Err_{NIR}=0.08$, which
indicates the intrinsic dispersion is $\sigma_{INT,
NIR}=\sqrt{\sigma_{NIR}^2-Err_{NIR}^2}=0.36$. The intrinsic
dispersion of the SED shape is substantial.

The E94 sources, which were corrected for host galaxy contribution,
are, by construction, clustered around the E94 mean in the
AGN-dominated region located at the bottom right corner of the
mixing diagram. For the 42 E94 sources, the mean slopes are
$\bar{\alpha}_{OPT}=0.96$ (standard deviation $\sigma_{OPT}=0.27$)
and $\bar{\alpha}_{NIR}=-0.50$ (standard deviation
$\sigma_{NIR}=0.38$). The mean of the measurement error of
$\alpha_{OPT}$ is $Err_{OPT}=0.1$, which indicates the intrinsic
dispersion is $\sigma_{INT,
OPT}=\sqrt{\sigma_{OPT}^2-Err_{OPT}^2}=0.25$. The mean of the
measurement error of $\alpha_{NIR}$ is $Err_{NIR}=0.2$, which
indicates the intrinsic dispersion is $\sigma_{INT,
NIR}=\sqrt{\sigma_{NIR}^2-Err_{NIR}^2}=0.32$.

To compare the intrinsic dispersion of these two samples, we
performed an F-test, ignoring the effects of the measurement error
and assuming the intrinsic slope distribution is normally
distributed. For $\alpha_{OPT}$, the F statistic is 1.09, and the
cumulative distribution function (CDF) probablity is 65\%, implying
that the difference between the $\alpha_{OPT}$ intrinsic dispersions
of R06 and E94 is not significant. For $\alpha_{NIR}$, the F
statistic is 1.13, the CDF probability is 66\%, implying the
difference between the $\alpha_{NIR}$ intrinsic dispersions of R06
and E94 is not significant. A more detailed statistical comparison
between the intrinsic dispersion in different samples is reported in
Hao et al. (2011).

\subsection{Selection of the HDP quasars}

In the mixing diagram, we define a circle with a radius of 0.6 to
approximate the 1.5$\sigma$ intrinsic dispersion region of the E94
mean SED template. We then use this region to classify outliers. As
the distribution of the quasars is continuous in $\alpha_{NIR}$,
different sized circles define different outlier populations. Here
we simply apply the same criterion as in Hao et al. (2010) to enable
ready comparison of their properties.

The Hao et al. (2010) mixing diagram readily distinguishes among the
AGN-dominated, galaxy-dominated and reddening-dominated SEDs. The
AGNs with SED slopes in the triangular region defined by the mixing
curve and the reddening vector can be explained by the combinations
of the E94 mean SED, host galaxy contamination and reddening. AGNs
which lie at least $1\sigma$ above the mixing curve and the
dispersion circle in the upper right corner of the mixing diagram,
are defined as HDP quasars. Note that we do not suggest that HDP
quasars are a distinct class. As the distribution of the quasars on
the mixing diagram is continuous and the definition of the
dispersion circle is somewhat arbitrary, the HDP quasars represent
one extreme of the distribution of the AGN SEDs.

We further divide the HDP quasars into three classes by their
positions relative to the equal slope line as in Hao et al. (2010).
Class I sources (blue points) have $\alpha_{OPT}>\alpha_{NIR}$
within measurement error. These sources have a normal blue bump but
flat NIR emission. Class II sources (green points) lie on the equal
slope line, which indicates that the NIR emission might be the
continuation of the optical accretion disk, although this turns out
to be not true for all cases judging from SED fitting in the
XMM-COSMOS sample (Hao et al. 2010).

We find 17 HDP quasars in the R06 sample. Most (15) of which belong
to Class I, with the remainder (2) belonging to Class II. The
fraction of the HDP quasars in R06 is $8.7\%\pm2.2\%$. For 3/4 of
these HDP quasars, more than 3 photometry points are used in the
near-infrared for slope estimation. The two Class II HDP quasars,
however, have only 3 NIR data points in their estimates.

Four sources out of the 42 quasars in the E94 sample qualify as HDP
(Table \ref{t:e94hdp}): 2 optically selected and 2 radio selected.
Their SEDs are shown in Figure \ref{e94sed}. They are all Class I
HDP quasars.

\section{HDP quasar properties}
\subsection{HDP Quasars in R06}
\label{s:r06hdp}
\begin{deluxetable}{ccccc}
\tabletypesize{\scriptsize} \tablecaption{HDP Quasars in R06
\label{t:r06hdp}} \tablehead{\colhead {Name (SDSS J)}
 & \colhead{redshift} & \colhead{log$L_{bol}$} & \colhead{Class} & \colhead{$N_{NIR}$\tablenotemark{a}}\\
 & & \colhead {erg$\cdot$s$^{-1}$} & & }\startdata
  105308.24+590522.2 & 0.430 & 45.16 &  I & 4\\
  160341.44+541501.5 & 0.581 & 45.27 &  I & 4\\
  104857.92+560112.3 & 0.800 & 45.70 &  I & 3\\\hline
  103525.05+580335.6 & 0.964 & 46.02 &  I & 4\\
  163306.12+401747.4 & 0.974 & 45.87 &  I & 4\\
  160913.18+535429.6 & 0.992 & 46.32 &  I & 4\\
  104935.76+554950.5 & 1.056 & 46.26 & II & 3\tablenotemark{b}\\
  104114.18+590219.4 & 1.093 & 45.93 &  I & 4\\
  104921.49+575036.6 & 1.106 & 46.08 &  I & 4\\\hline
  171441.11+601342.9 & 1.456 & 46.05 &  I & 4\\
  104948.86+592620.7 & 2.000 & 46.22 &  I & 4\\
  103905.24+591209.1 & 2.115 & 46.50 &  I & 4\\
  104226.20+585925.8 & 2.560 & 46.48 &  I & 4\\
  160828.33+535251.9 & 2.960 & 46.61 & II & 3\\
  164238.08+412104.7 & 3.123 & 46.76 &  I & 4\\
  104810.81+575526.9 & 3.171 & 46.71 &  I & 4\\
  105654.96+583712.4 & 3.856 & 46.81 &  I & 3\\

\enddata
\tablenotetext{a}{Number of photometry points used in the
$\alpha_{NIR}$ estimates.} \tablenotetext{b}{$\alpha_{NIR}$
 may be overestimated (see section~\ref{s:robust}, Figure~\ref{dz}).
 Detailed discussion about this source is in section~\ref{s:r06hdp}.}
\end{deluxetable}

\begin{figure*}
\includegraphics[angle=0,width=0.48\textwidth]{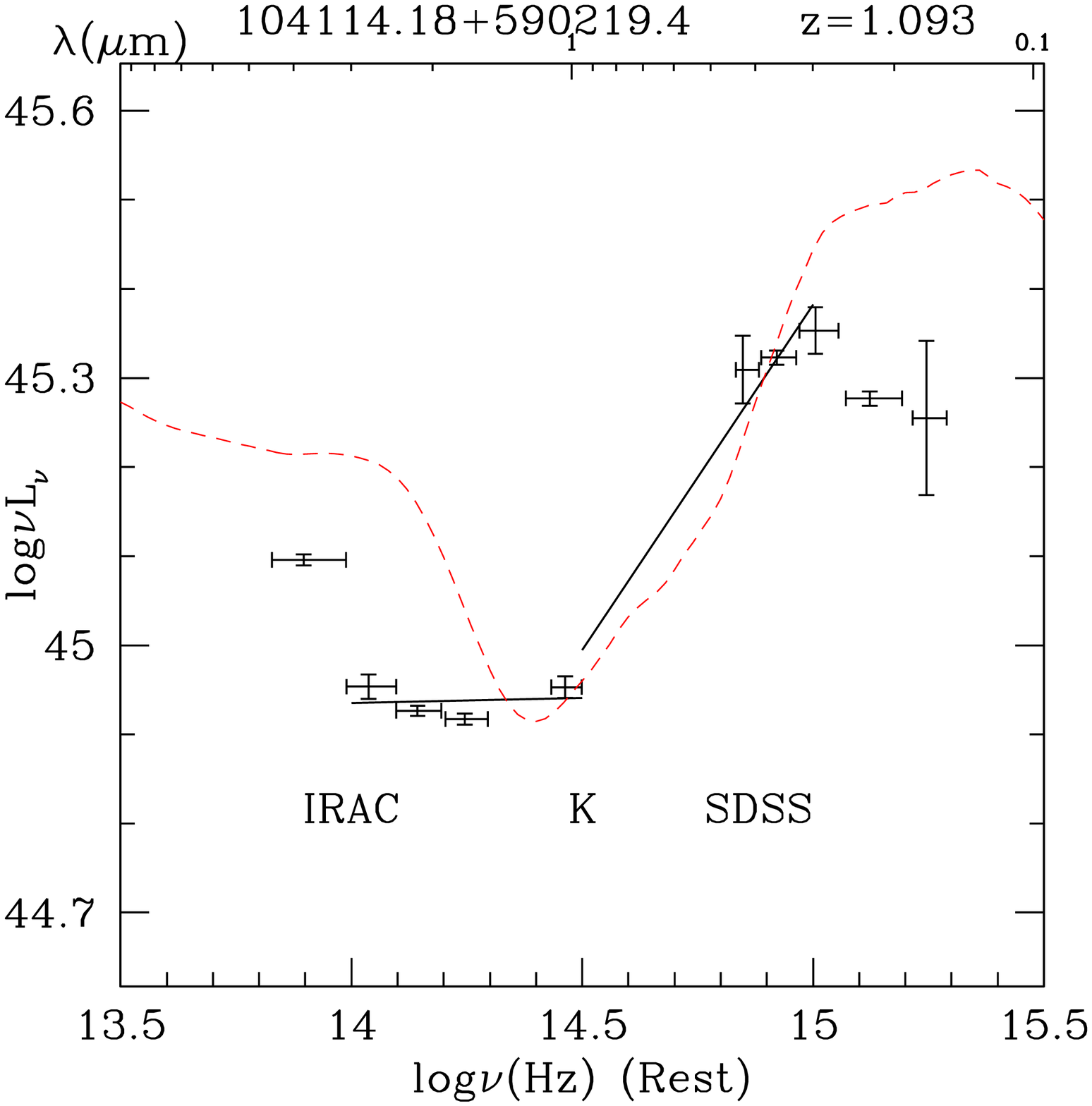}
\includegraphics[angle=0,width=0.48\textwidth]{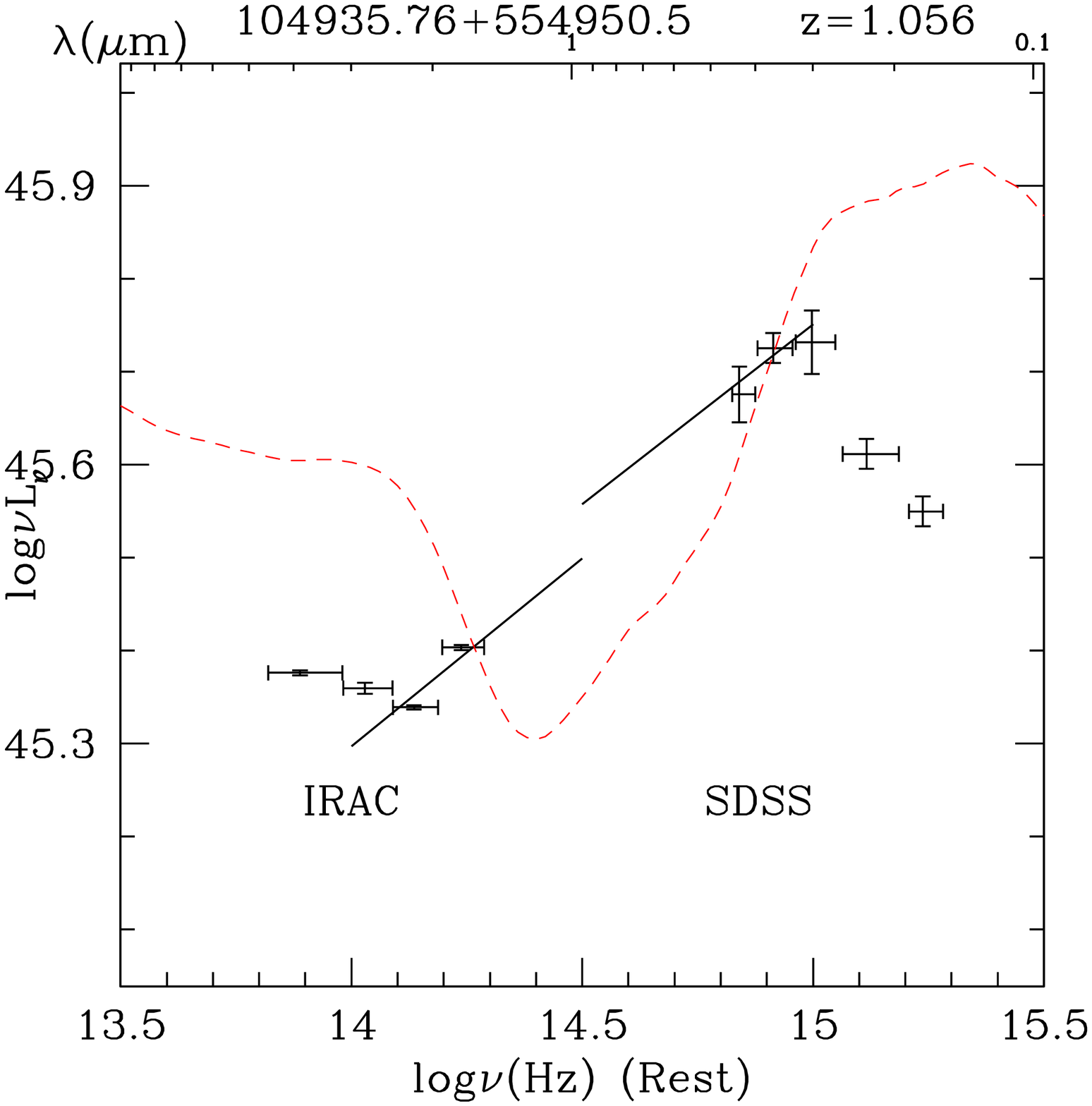}
\caption{Examples of HDP SEDs in R06: {\em Left:} Class I; {\em
Right:} Class II ($\alpha_{OPT}$ and $\alpha_{NIR}$ less reliable
see section~\ref{s:robust}). The red dashed line is the E94 radio
quiet mean SED.\label{egSED}}
\end{figure*}

\begin{figure}
\includegraphics[angle=0,width=0.48\textwidth]{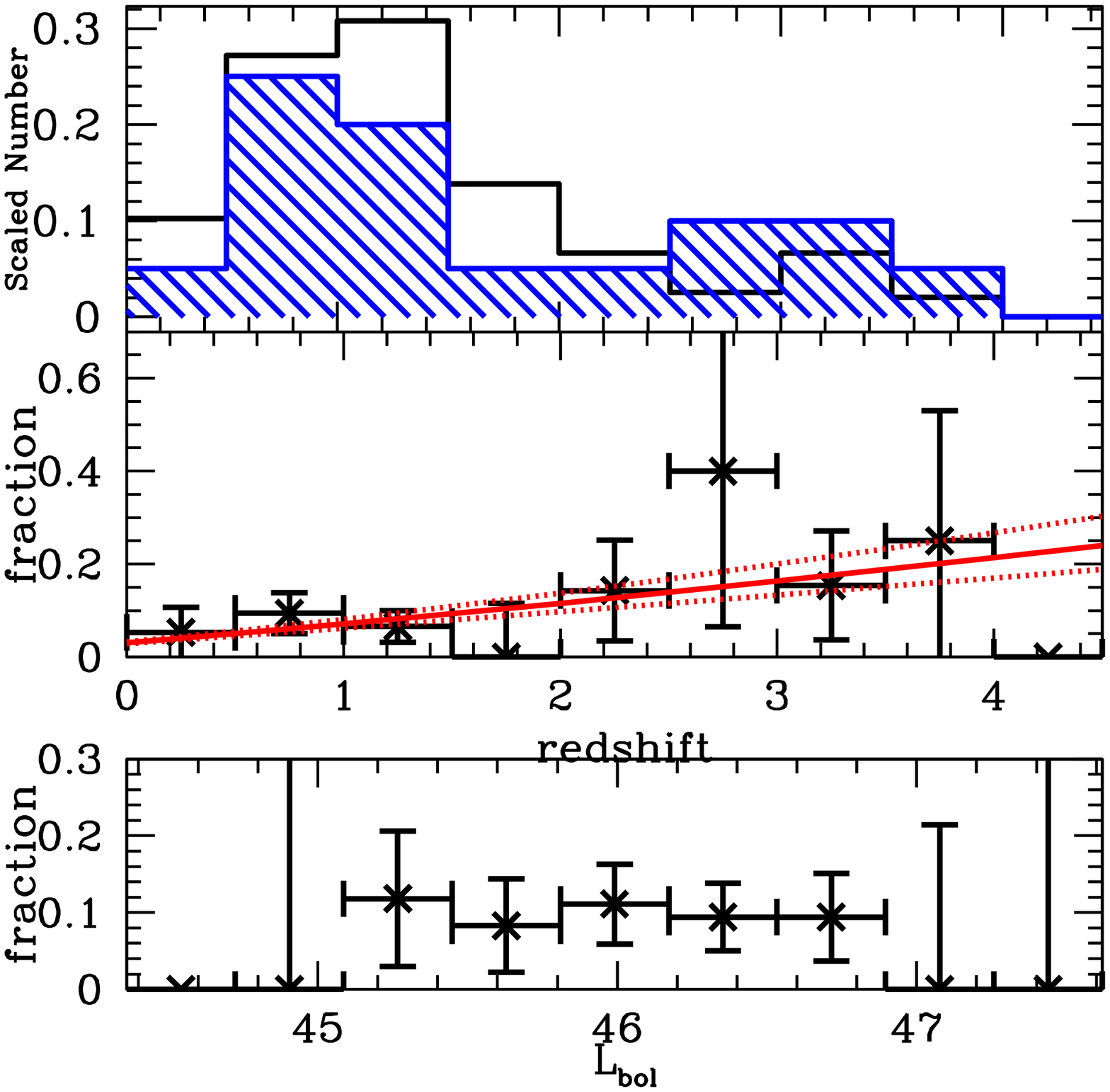}
\caption{The redshift distribution (top) and fraction of the HDP AGN
in R06 sample as function of redshift (middle) and $L_{bol}$
(bottom). In the top panel, the black histogram shows the redshift
distribution of the R06 sample. The blue shaded histogram shows the
distribution of the HDP quasar in R06 sample. The two histogram are
scaled to have the same integral area. In the middle panel, the red
solid line shows the fitting of the XMM-COSMOS HDP fraction versus
redshift in the same bin
($f_{HDP}=(0.031\pm0.002)(1+z)^{(1.2\pm0.1)}$) and the red dashed
line show the 1$\sigma$ range of the fitting. \label{frac}}
\end{figure}

Representative R06 SEDs (Class I: top, Class II: bottom) are shown
in Figure~\ref{egSED}. The Class I source (SDSS J104114.18+590219.4)
by definition show a flat near infrared SED, similar to the Class I
HDP quasars found in XMM-COSMOS quasar sample (Hao et al. 2010).

The Class II source (SDSS J104935.76+554950.5) has only 3 points in
both slope estimates and is at $z=1.056$, in the range where
$\alpha_{NIR}$ can be overestimated by up to $\sim 0.5$ (Figure
\ref{dz} in section~\ref{s:robust}). However it is hard to fit the
SED with an E94 mean SED and some host. The NIR SED of this source
shows an unusual discontinuous, 0.1 dex arise from the IRAC2 to
IRAC1 points at $\sim2\mu m$ (rest). This feature cannot be caused
by the emission from hot dust, because a black-body, at any
temperature below the maximum dust sublimation temperature of 1900
K, does not fit the feature well. The Paschen $\alpha$ line lies in
the IRAC 1 band but is unlikely to produce this feature, as to
produce the IRAC1 rise would require an equivalent width of
$\sim$2000\AA, which is 30 times the typical value of 70\AA\ (Landt
et al. 2008). As $\alpha_{OPT}$ is within the range of those for E94
quasars (section~\ref{s:e94hdp}), it could be that the IRAC1 point
is dominated by the continuation of the accretion disk. NIR
photometry or spectra are needed to understand the emission in this
range.

As reported in Hao et al. (2010), the fraction of the HDP quasars in
the XMM-COSMOS sample evolves with redshift but not with other
parameters (e.g., $L_{bol}$). The fraction increases from 6\% at low
redshift ($z<2$) to 20\% at moderate high redshift ($2<z<3.5$). We
checked if similar evolution behavior can be found in the R06
sample. We plotted the HDP fraction in R06 versus $z$ and $L_{bol}$
in Figure \ref{frac}. The HDP quasars in the R06 sample have
redshifts from 0.43 to 3.86. The range of $L_{bol}$ for the HDP AGNs
(45.1$<$log$L_{bol}<$46.8) is almost the same as for the whole
sample (44.8$<$log$L_{bol}<$47.4). The Kolmogorov-Smirnov (KS) test
shows that the distributions of z and $L_{bol}$ for the R06 HDP
quasars are indistinguishable from the rest of the R06 sample (KS
probabilities of 0.25 and 0.88). As the XMM-COSMOS sample, the HDP
quasars in R06 do not show a different $L_{bol}$ distribution to the
other quasars in R06. However, the HDP quasars in R06 do not show
redshift evolution, unlike XMM-COSMOS sample. This result might be
affected by the 64 sources we have excluded due to limited
photometry coverage. At redshifts $z<0.5$, the 5 sources with no J H
K photometry have fewer than three photometry points in the rest
frame $1\mu m - 3\mu m$ band, as do the 6 quasars at high redshifts
$z>3.7$. At redshifts $1.5<z\leqslant3.7$, 53 quasars also have
fewer than three photometry points in the rest frame $0.3-1\mu m$.
So quasars at low and high redshifts have more chance of being
excluded from consideration, which might lead to the lack of an
increase in HDP fraction in R06 at $z\sim 2$, though $\alpha_{OPT}$
only has a small influence on HDP definition. The XMM-COSMOS sample
has no such incompleteness bias, since the COSMOS photometry
coverage is more complete and uniform.

In the XMM-COSMOS sample, the HDP fraction jumps at $z=2$ from $\sim
6\%$ to $\sim 20\%$. So we re-binned the R06 sample and find that
the HDP fraction is $6.3\%\pm2.1\%$ at $z<2$, and $19.4\%\pm8.0\%$
at $z\geqslant2$ respectively, a $1.6\sigma$ difference, which is
consistent with the results in the XMM-COSMOS sample. We over-plot
XMM-COSMOS fit of the HDP fraction versus $z$ (red lines,
$f_{HDP}=(0.031\pm0.002)(1+z)^{(1.2\pm0.1)}$, Hao et al. 2010). The
HDP fraction versus $z$ relationship in the two samples do not
disagree.

Because of the poorly sampled SED at $\sim 1 \mu m$ due to limited
NIR photometry in the R06 sample, we could not make robust estimates
of the dust covering factor ($f_c$) or the outer edge of the
accretion disk ($r_{out}$).

\subsection{HDP Quasars in E94}
\label{s:e94hdp}
\begin{deluxetable}{lclcccccccccc}
\tabletypesize{\scriptsize} \tablecaption{HDP Quasars in E94
\label{t:e94hdp}} \tablehead{\colhead{Object} & \colhead {Name}
 & \colhead{redshift} & \colhead{log$L_{bol}$} & \colhead{log($M_{BH}$)}
 & \colhead{$\lambda_E$} & \colhead {$T_d$} & \colhead{$A_d$} & \colhead {$f_c$}
 & \colhead{$T_{out}$} & \colhead {$R_{out}$} & \colhead {$R_{out}/$} & \colhead {$R_{out}/$}\\
 & & & \colhead {erg$\cdot$s$^{-1}$} & \colhead {$M_{\bigodot}$}&
 & \colhead {K} &\colhead{pc$^2$} & \colhead{\%} & \colhead{K}
 & \colhead{pc} &\colhead {$r_s$} & \colhead {$r_{grav}$}}\startdata
Q0003+158 & PHL 658 & 0.45 & 46.5 & 9.3\tablenotemark{1} & 0.13 & 1200 & 8.00 & 6 & 3700 & 0.6 & 3200 & 11 \\
Q0049+171 & PG 0049+171 & 0.064 & 44.8 & 8.3\tablenotemark{1} & 0.03 & 1300 & 0.13 & 11 & 4000 & 0.06 & 2700 & 11 \\
Q0414-060 & 3C110 & 0.78 & 47.0 & 9.9\tablenotemark{2} & 0.10 & 1900 & 3.80 & 18 & 3600 & 0.4 & 370 & 13 \\
Q1635+119 & MC2 1635+119 & 0.146 & 45.4 & 8.1\tablenotemark{3} & 0.16 & 1900 & 0.15 & 21 & 4800 & 0.08 & 6300 & 9 \\
\enddata
\tablenotetext{1}{Reference: Vestergaard \& Peterson (2006)}
\tablenotetext{2}{Reference: Marziani et al. (2010)}
\tablenotetext{3}{Reference: Woo et al. (2002)}
\end{deluxetable}

\begin{figure*}
\includegraphics[angle=0,width=0.24\textwidth]{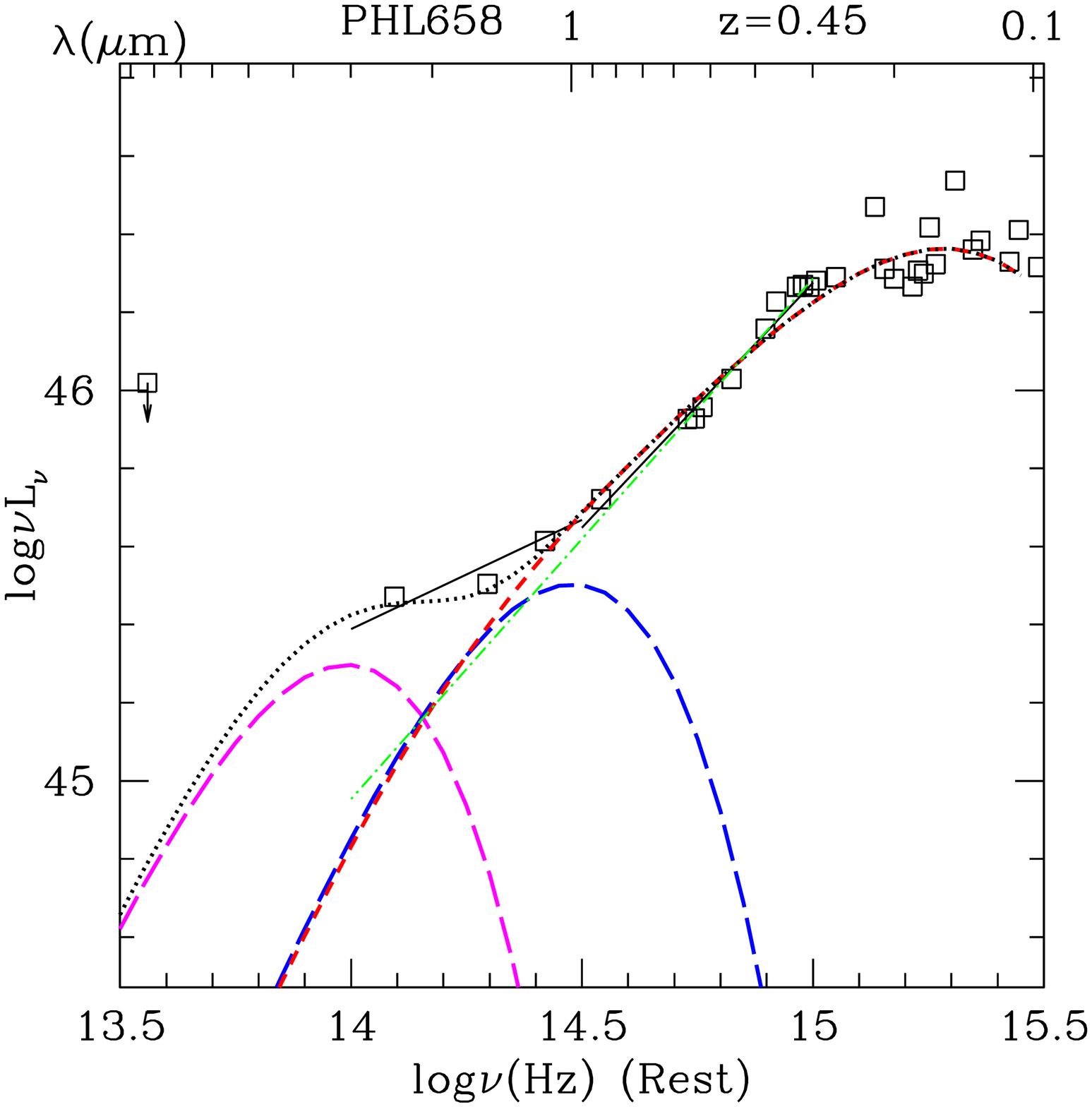}
\includegraphics[angle=0,width=0.24\textwidth]{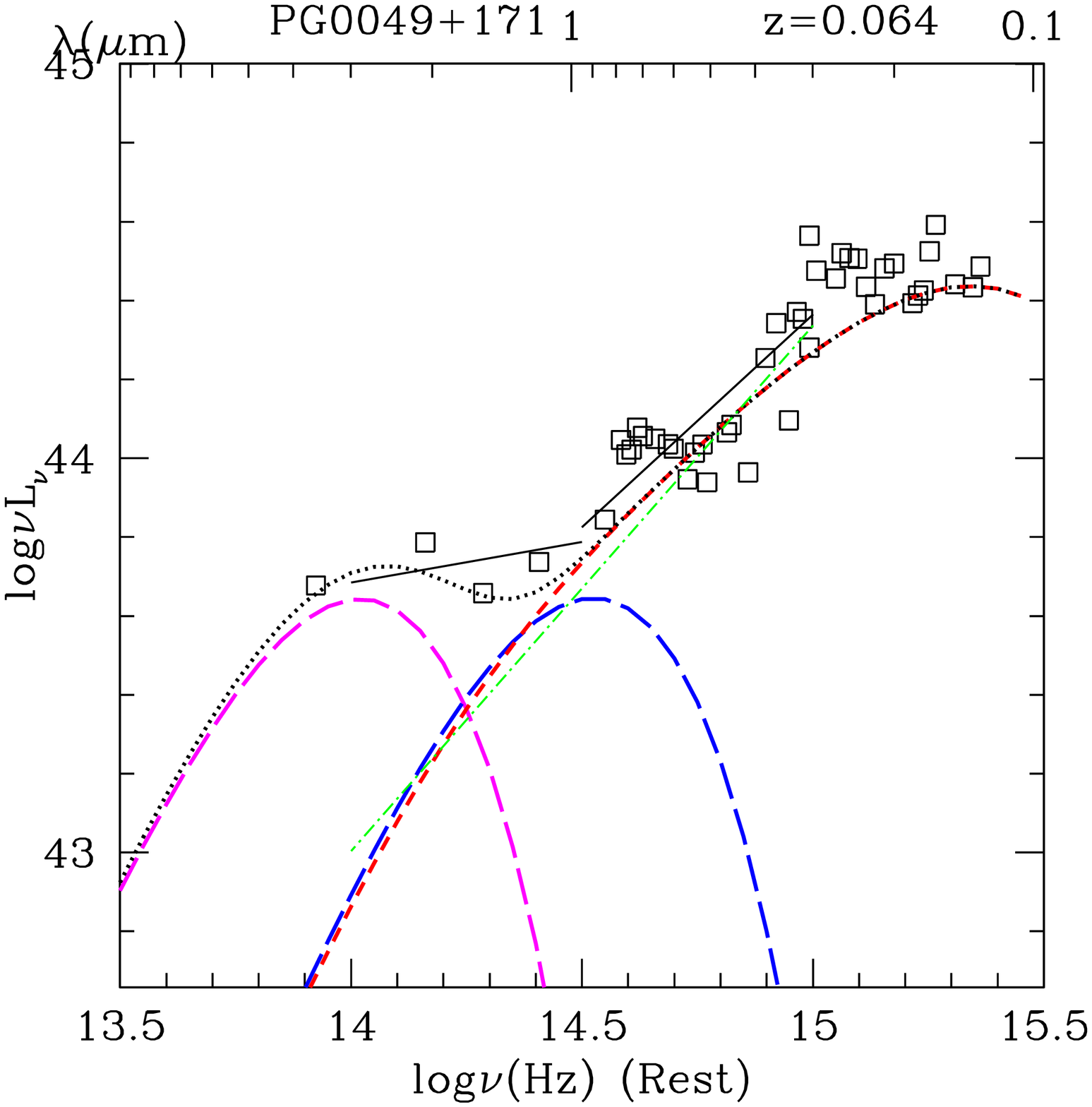}
\includegraphics[angle=0,width=0.24\textwidth]{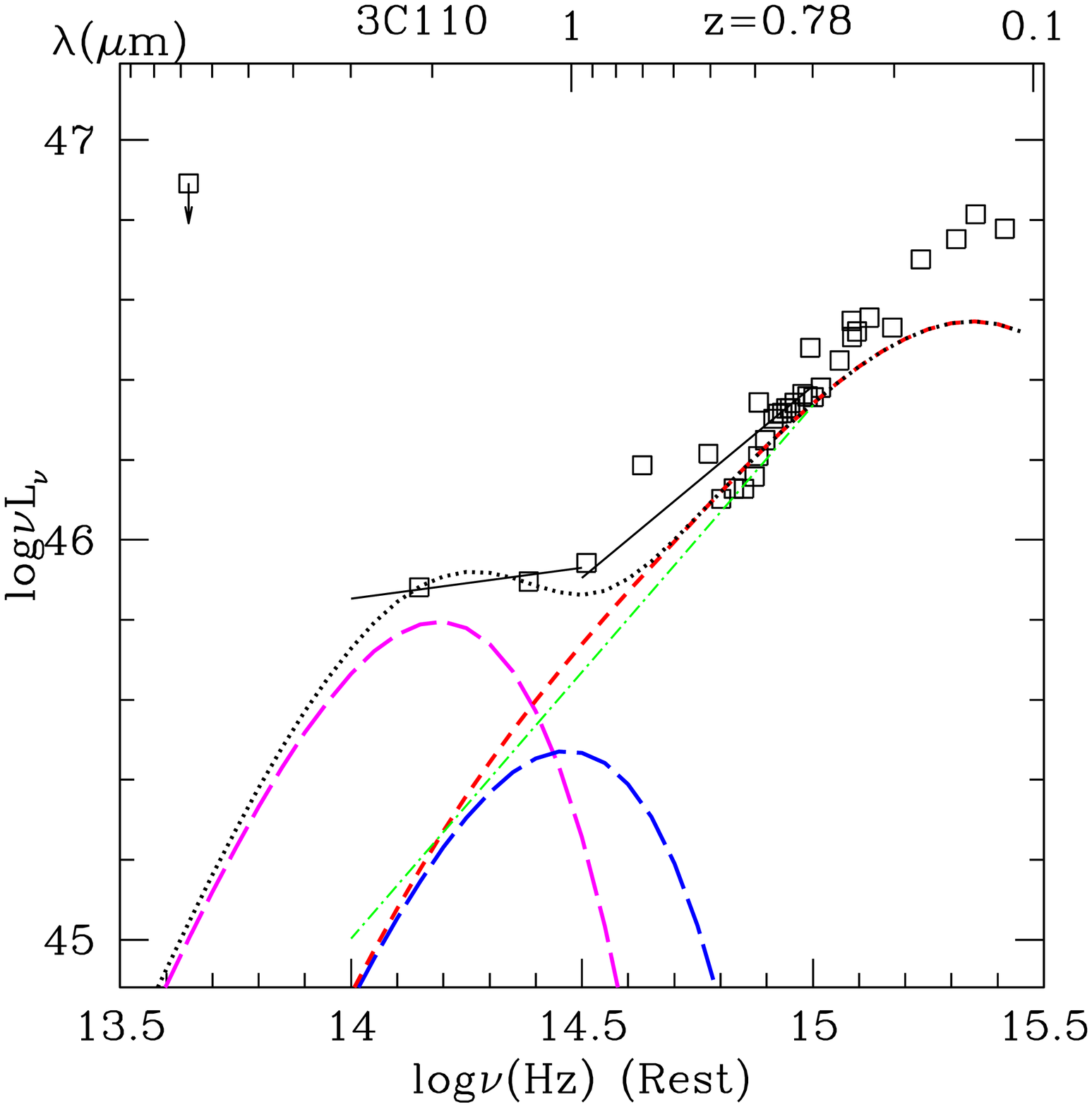}
\includegraphics[angle=0,width=0.24\textwidth]{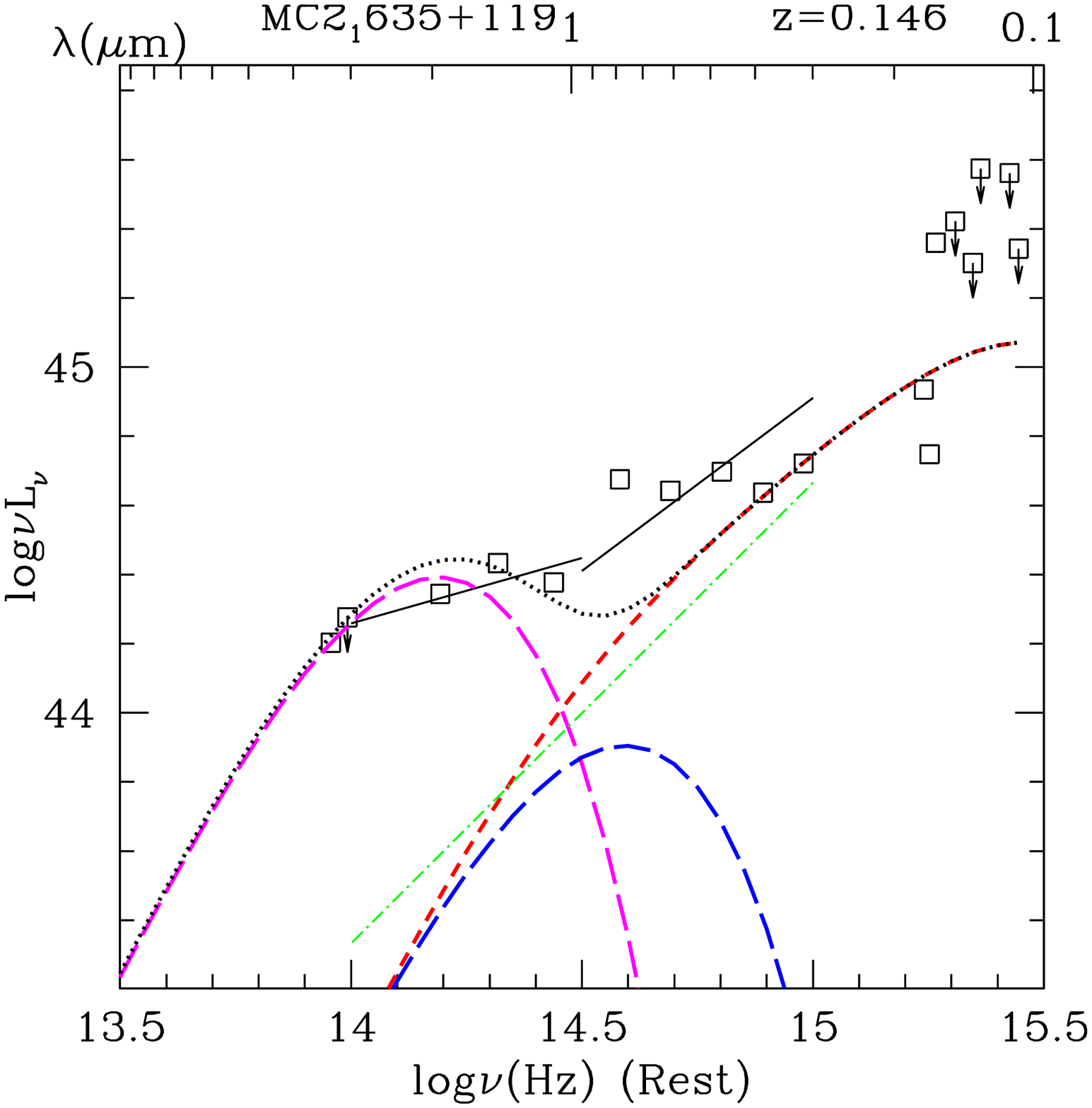}
\caption{HDP AGN in E94 sample: {\em Left:} Q0003+158 (PHL 658,
redshift z=0.45); {\em Middle left:} Q0049+171 (PG0049+171,
z=0.064); {\em Middle right:} Q0414-060(3C110, z=0.78); {\em Right:}
Q1635+119(MC2 1635+119, z=0.146). The SEDs are fitted with the
accretion disk component (red dashed line) and a hot dust component
(magenta dashed line). The sum of the two component is showed as
black dotted line. The blue dashed line (a single temperature black
body) shows the fitting of the outer edge of the accretion disk. The
green dot-dashed line is the spectrum from a steady optically thick
accretion disk ($\nu F_\nu \varpropto \nu^{4/3}$). \label{e94sed}}
\end{figure*}

The fraction of HDP AGNs in E94 is $9.5\%\pm 5.0\%$. Because the
sample size of E94 is 10 times smaller than that of the XMM-COSMOS
type 1 AGN sample, the error on the fraction is much larger. A KS
test shows that the distributions of $z$ and $L_{bol}$ for these HDP
quasars are indistinguishable from the rest of the E94 sample (with
KS probabilities of 0.5 and 0.6 respectively). These results agrees
with the XMM-COSMOS sample that for low redshift quasars, the
fraction of the HDP do not evolve with $z$ and $L_{bol}$.

As the E94 SEDs are all corrected for host galaxy contributions, we
fit the SED with just two components: an accretion disk (using a
standard Schwarzschild $\alpha$-disk model, with electron scattering
and Comptonization; Siemiginowska et al. 1995), and a hot dust
component (using a single temperature black-body). The black hole
masses for these four quasars, reported in Vestergaard \& Peterson
(2006), Marziani et al. (2010) and Woo et al. (2002), is listed in
Table \ref{t:e94hdp}. Their accretion rates (expressed as their
Eddington ratios, $\lambda_E$) are calculated from $L_{bol}$ and
$M_{BH}$. We use the accretion disk model (Siemiginowska et al.
1995) for the particular black hole masses and the accretion rates
for each quasar and show the results in Figure \ref{e94sed}.

\subsubsection{Covering Factors of E94 HDP Quasars}
Following the same method as in Hao et al. (2010), the dust covering
factor ($f_c$) was estimated as the ratio of the dust emission area
($A_d$) to the area at the dust evaporation radius ($A_e=4\pi
r_e^2$). $A_d$ comes from the normalization of the black-body fit.
The dust evaporation radius ($r_e$) is estimated from $r_e=1.3L_{uv,
46}^{1/2}T_{1500}^{-2.8} pc$ (Barvainis 1987), where $L_{uv,46}$ is
the total ultraviolet (1$\mu$m--912\AA) luminosity in units of
$10^{46}$ erg s$^{-1}$, and $T_{1500}$ is the maximum dust
temperature allowed by the SED, in units of 1500~K. The typical
errors on the E94 photometry in the near-infrared are $\sim 0.1$
dex. This produces an error on the dust temperature estimation,
$\Delta T_d/T_d\sim8\%$, and an error on the normalization, $\Delta
A_d/A_d\sim4\%$. The error on $f_c$ is thus $\Delta f_c/f_c\sim5\%$.

\subsubsection{Accretion Disk Outer Radii of E94 HDP Quasars}

The outer accretion disk radius is estimated from the standard
$\alpha$--disk model formula (Frank, King \& Raine, 2002)
$$R_{out}=1.1\times 10^4 T_c^{-\frac43}\alpha^{-\frac{4}{15}}
\eta^{-\frac25}M_8^{\frac{11}{15}}\lambda_E^{\frac25}f^{\frac85}~~~pc$$
where $M_8=M/(10^8 M_{\bigodot})$, $\lambda_E=L_{bol}/(\frac{4\pi
Gcm_p}{\sigma_e}M)$, and $f=\left [1-\left (\frac{6GM}{R~c^2}\right
)^{\frac{1}{2}}\right ]^{\frac{1}{4}}$. We assume $\alpha=0.1$,
$\eta=0.1$. Thus $R_{out,pc}\propto
T_{out}^{-\frac43}M^{\frac{11}{15}}\lambda_E^{\frac25}\propto
T_{out}^{-\frac43}M^{\frac13}$.

The error in the disk temperature estimates from the E94 photometric
errors is $\Delta T_{out}/T_{out}\sim5\%$. Unfortunately, black hole
mass estimates from mass scaling relationships have an error $\Delta
M_{BH}/M_{BH}\sim40\%$ (Vestergaard \& Peterson 2006, Peterson
2010), so $\Delta R_{out, pc}/R_{out, pc}\sim 20\%$. In terms of
Schwarzschild radii ($r_s=2GM/c^2$), $\Delta R_{out, r_s}/R_{out,
r_s} \sim 33\%$.

We can also express this radius in terms of the gravitational
instability radius of the accretion disk ($r_{grav}$, Goodman 2003),
which is the radius beyond which the disk is unstable to
self-gravity and should break up. As $r_{grav}\propto
M^{-\frac29}\lambda_E^{\frac49}\propto M^{-\frac23}$, $R_{out,
r_{grav}}\propto T_{out}^{-\frac43}$. Hence, $\Delta R_{out,
r_{grav}}/R_{out, r_{grav}}\sim 7\%$.

The results are reported in Table \ref{t:e94hdp}. They turn out to
be comparable to the HDP quasars in the XMM-COSMOS sample in that
$R_{out} \sim 10 r_{grav}$, a surprisingly large value.

\section{Discussion and Conclusions}
\subsection{Comparison with COSMOS HDP Quasars}
We find a similar fraction of HDP quasars in the optically selected
R06 and E94 samples as in the X-ray selected XMM-COSMOS sample of
Hao et al. (2010): $8.7\%\pm2.2\%$ in the R06 sample and $9.5\%\pm
5.0\%$ in E94 sample. This similarity indicates that these extreme
AGNs are fairly common even in optically selected samples, though
they were not previously recognized.

The XMM-COSMOS quasar sample and the R06 sample have similar
redshift coverage. In contrast to the XMM-COSMOS sample, however,
the fraction of the HDP in R06 does not show significant evolution
with redshift according to the KS test result. This difference might
be caused by the exclusion of the 64 sources in certain redshift
range due to the limit near-IR photometry. The evolution of the HDP
fraction in R06 sample is ill-determined but generally agrees with
the HDP of XMM-COSMOS sample (Figure \ref{frac}).

In the small E94 sample at low $z$, the redshifts and luminosities
of the HDP quasars are indistinguishable from the normal type 1
AGNs. This is consistent with the XMM-COSMOS sample, as for low
redshift sources ($z<2$), the HDP fraction does not evolve with
redshift and luminosity.

All of the HDP quasars in R06 are Class I and II. In comparison, the
XMM-COSMOS sample has 27\% (11 out of 41) Class III HDP quasars. We
compare the Class III distribution versus Class I and II in the R06
HDP and XMM-COSMOS HDP samples using the Fisher exact test (Wall \&
Jenkins 2003). The probability that the two HDP samples have the
same class distribution is 1.4\%. The SEDs of the Class III HDP
quasars have stronger galaxy contribution than Class I and II. So
R06 includes fewer sources with strong galaxy contribution than
XMM-COSMOS. This is expected, as the R06 sample is selected by
optical colors, and will exclude low AGN-to-galaxy contrast objects.
The XMM-COSMOS sample, instead, is X-ray selected, and includes
sources with strong galaxy contributions.

In E94, the dust covering factor of the HDP quasars ranges from 6\%
to 21\%, similar to the XMM-COSMOS HDP quasars (Hao et al. 2010). A
small dust covering factor was proposed long ago in several quasars.
For example, the spectropolarimetry analysis of the quasar OI 287
showed that the broad Balmer lines are polarized the same way as the
continuum, while the forbidden lines have zero intrinsic
polarization, suggesting a thin `torus' (Goodrich \& Miller, 1988).

The outer edge of the accretion disk in E94 (Table~\ref{t:e94hdp})
ranges from 0.06pc to 0.6pc and from 370$r_s$ to 6300$r_s$, which
are $\sim$ 10 times greater than the gravitational instability
radius (Goodman 2003). A similar extension of the disk spectrum into
the NIR uncovered by polarized light, was found by Kishimoto et al.
(2008). The outer radius of the accretion disk is further out than
expected by simple $\alpha$-disk theory. For the R06 sample, the
limitations of the photometry prevents the accurate estimation of
these quantities.

\begin{figure}
\includegraphics[angle=0,width=0.48\textwidth]{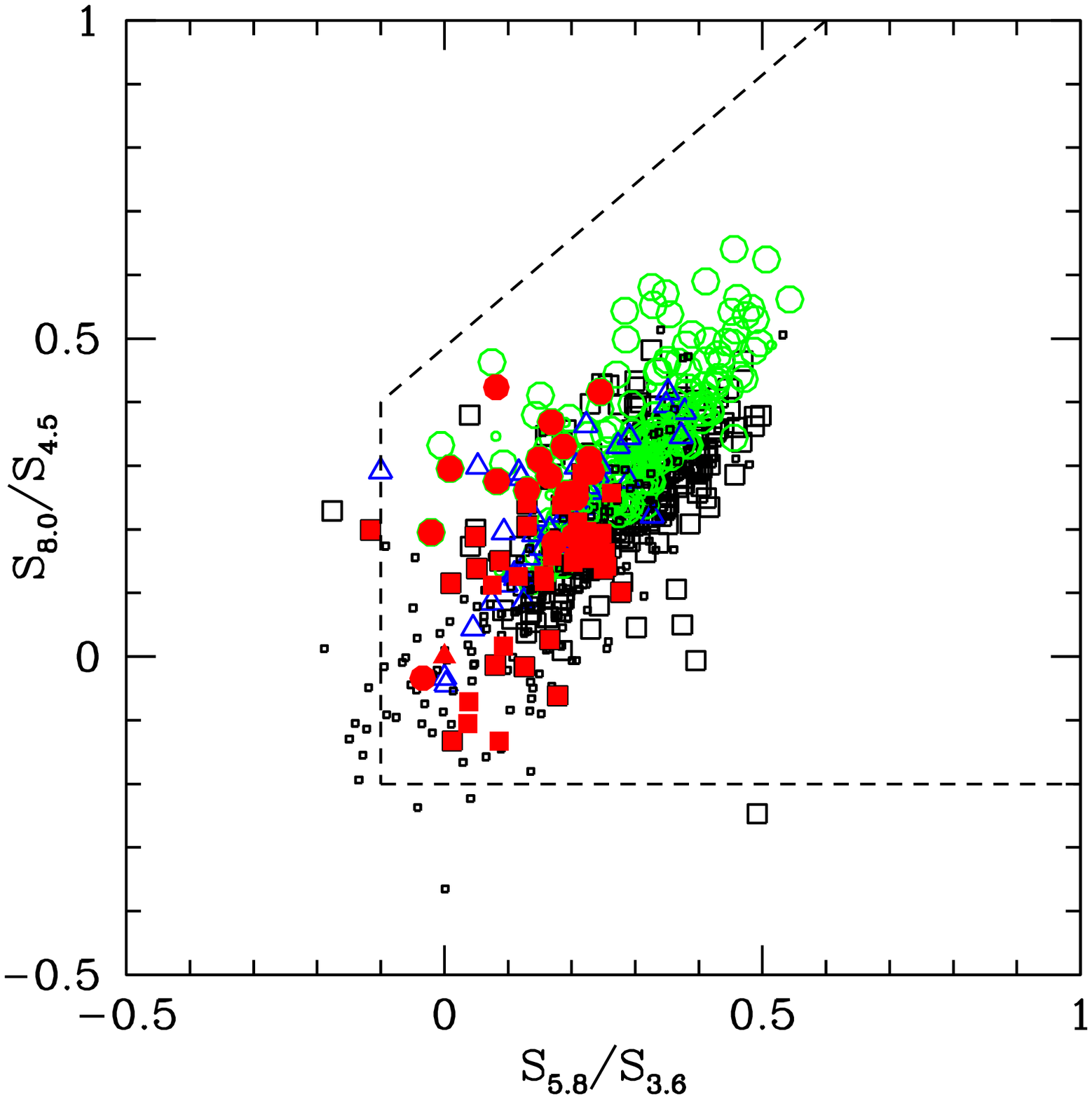}
\caption{Position in Lacy et al. (2004) color-space of the
XMM-COSMOS (square), R06 (circle) and E94(triangle) samples. The red
solid points are the HDP quasars from these samples. Small sets of
symbols are for sources with $\alpha_{OPT}<0.2$, that is galaxy- or
reddening-dominated sources. \label{lacy}}
\end{figure}

\begin{figure*}
\includegraphics[angle=0,width=0.48\textwidth]{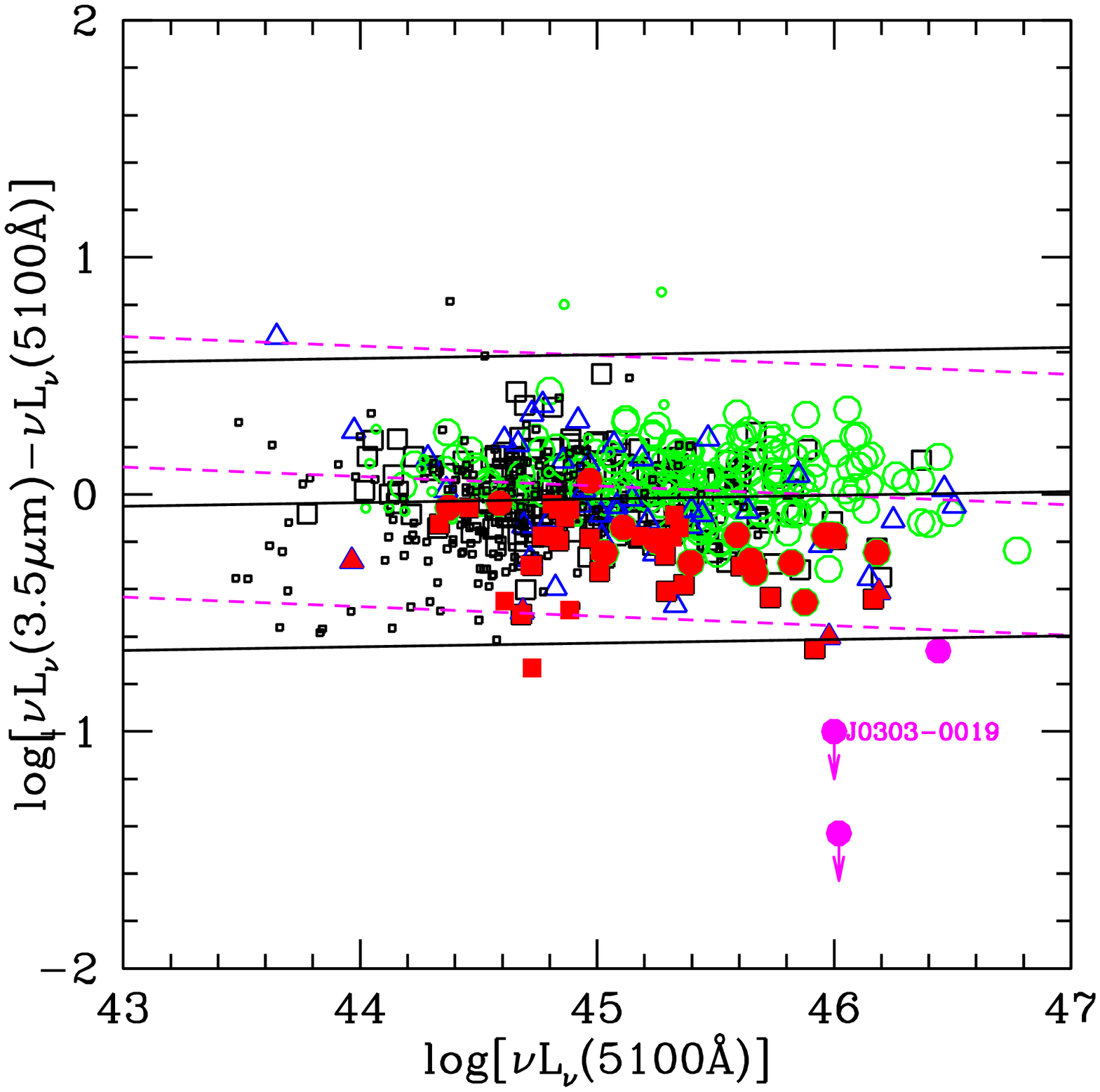}
\includegraphics[angle=0,width=0.48\textwidth]{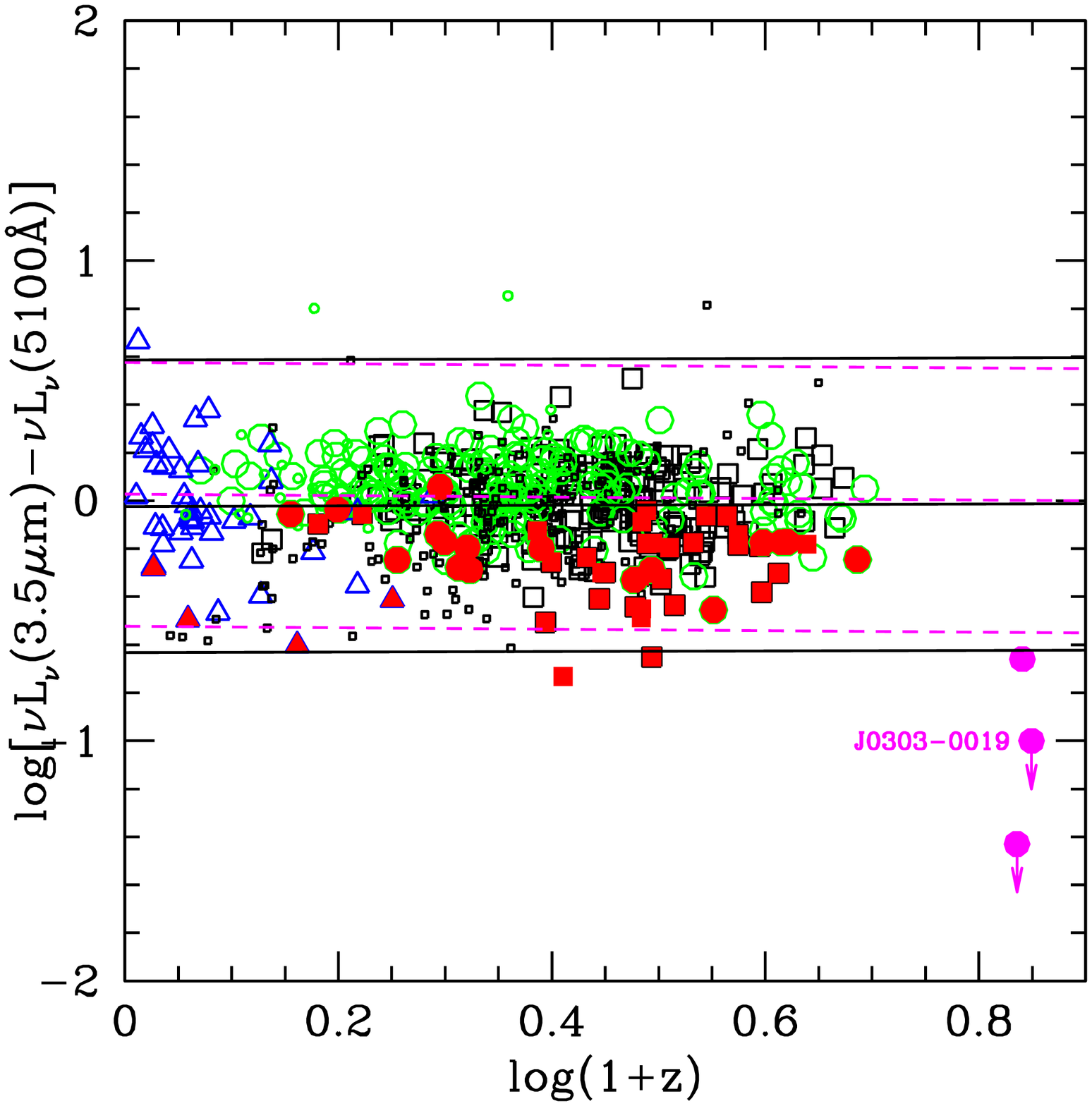}
\caption{Luminosity and redshift dependence of the hot-dust
abundance introduced in Jiang et al. (2010) of the XMM-COSMOS
(square), R06 (circle) and E94(triangle) samples. The red solid
points are the HDP quasars from these samples. The magenta points
are the three $z\sim6$ quasars in Figure 1 of Jiang et al. (2010)
that is HDP. Small sets of symbols are for sources with
$\alpha_{OPT}<0.2$, that is galaxy- or reddening-dominated sources.
The magenta dashed lines are the linear fit and $3\sigma$ range of
the sample in Jiang et al. (2010). The black solid lines are the
linear fit and $3\sigma$ range of the XMM-COSMOS, R06 and E94
samples. \label{jiang}}
\end{figure*}

\subsection{HDP Quasars and IR Quasar Selection}

We checked whether the HDP quasars lie within the Lacy et al. (2004)
color-selection region for quasars (Figure \ref{lacy}). Most of the
HDP quasars lie in the bottom left of the quasar region. Although
their hot dust emission is weak, it is still sufficient to
differentiate them from starburst galaxies. The HDP quasars are one
of the extreme of the variety of the AGN SEDs. AGNs with
galaxy-dominated SEDs lie in the same region. The Lacy et al. (2004)
color-color plot does not distinguish the HDP quasars from
galaxy-dominated quasars.

\subsection{Comparison with Jiang et al. Hot-Dust-Free Quasars}
Jiang et al. (2010) introduced the rest-frame 3.5$\mu$m to 5100\AA\
luminosity ratio as an indicator of the hot-dust abundance. They
concluded that there are no low redshift counterparts to the two
$z\sim 6$ hot-dust-free quasars J0005-0006 and J0303-0019. We
reproduce the Jiang et al. (2010) plot in Figure \ref{jiang} for the
XMM-COSMOS, R06 and E94 samples. The HDP quasars from COSMOS, R06
and E94 (red points) do lie at the bottom edge of the distribution,
but are not as extreme as the two Jiang et al. `hot-dust-free'
quasars. However, the $L(3.5\mu m/L(5100\AA)$ ratio alone cannot
distinguish the HDP from the galaxy -dominated SED shape (e.g.,
small black points), because they have similar rest frame 3.5 $\mu$m
to 5100\AA\ luminosity ratios.

We plot the 6 quasars from Jiang et al. (2010) on the mixing diagram
(Figure~\ref{md}, magenta dots). We find that three of the 6 Jiang
et al. (2010) $z\sim6$ quasars (J0005-0006, J0303-0019 and
J1411+1217) qualify as HDP quasars. This $\sim 50\%$ HDP fraction at
$z\sim6$ suggests, at low significance, that HDP quasars fraction
continue to grow with redshift. All three are class II, i.e. their
NIR emission may be a continuation of the accretion disk.
J1411+1217, which was classified as a normal quasar in Jiang et al.
(2010), is actually HDP by our criteria.

One difficulty with $\nu L_{\nu}(3.5\mu m)/\nu L_{\nu}(5100\AA)$ is
that this ratio does not allow for a range in optical slope, such as
would be induced by moderate reddening. The slope for the rest frame
optical SED of J0303-0019 ($\alpha_{OPT}\sim0.6$) is flatter than a
standard accretion disk, so that a continuation of the power law to
the rest frame infrared underestimates the hot dust contribution.
De-reddening the optical SED leads to a requirement for a host dust
excess. So this quasar should be counted as ``hot-dust-poor'' rather
than ``hot-dust-free''. Another difficulty with $\nu L_{\nu}(3.5\mu
m)/\nu L_{\nu}(5100\AA)$ is that sources with weak hot dust emission
but some cool dust emission are not selected by this parameter (e.g.
XMM-COSMOS HDP quasar XID=96, Hao et al. 2010). Defining an index
based on only two rest-frame wavelength band is not so effective at
low redshift. The mixing diagram of Figure~\ref{md} is more useful
in selecting the outliers.

\subsection{Origin of the HDP Quasars}
The origin of HDP quasars is unknown. As for the COSMOS HDP quasars
discussed in Hao et al. (2010), the E94 and R06 HDP quasars are at
$z<4$, when the universe is more than 2 Gyr old. They have more than
1 Gyr from reionization to form a torus. So they cannot be the first
generation of quasars. They are also luminous enough, and have high
enough accretion rates, to support a dusty torus (Elitzur \& Ho
2009).

Either the hot dust is destroyed (dynamically or by radiation), or
the dust is not centered on the SMBH, i.e., an off-nuclear AGN
(Blecha et al. 2011, 2008, Guedes et al. 2010, Volonteri \& Madau
2008). When an SMBH recoils (or is kicked-out), it is possible bring
along the adjacent broad line region, but not the dusty torus which,
being further out, is less tightly bound to the SMBH (Loeb 2007).
The dust is thus not centered on the SMBH, i.e. an off-nuclear AGN.
A good candidate for a recoiling BH has been found in the COSMOS
sample (Civano et al. 2010), although it is not included in the
XMM-COSMOS HDP sample due to the galaxy-dominated NIR--OPT SED
shape.

Volonteri \& Madau (2008) estimated the cumulative number of
off-nuclear AGNs (offset$>0.2$\asec) per square degree versus
redshift. Their lower limit on the number of off-nuclear AGNs is
$\sim2~deg^{-2}$ at $z<1$, $\sim8~deg^{-2}$ at $z<2$, and
$\sim11~deg^{-2}$ at $z<3$. Considering the COSMOS field is
$2~deg^2$, the Volonteri \& Madau (2008) prediction is consistent
with the cumulative number of XMM-COSMOS HDP AGNs ($2.5~deg^{-2}$ at
$z<1$, $8~deg^{-2}$ at $z<2$, $18.5~deg^{-2}$ at $z<3$). A detailed
comparison between the theoretical distribution and the HDP
evolution will be reported in a following paper. We could not make
similar estimates for the R06 and E94 samples, because the survey
areas are not well defined.

Alternatively, the existence of HDP quasars in samples with
different selection methods and the continuous distribution of dust
covering factor imply that the origin of HDP quasars could be
related to the AGN structure. Misaligned disks will result from
discrete isotropic accretion events (Volonteri et al. 2007), which
will lead to a wide range of covering factors (Lawrence \& Elvis
2010). For disks with tilt-only warps (i.e. with no rotation of the
line of nodes), $\sim 14\%$ of the type 1 AGNs will have covering
factors less than 20\% (Lawrence \& Elvis 2010). This agrees with
the HDP fraction in all three samples.

\subsection{Prospects}
The ongoing UKIDSS and WISE surveys will expand the available HDP
samples greatly, especially in the $z>1.5$ range, where evolution
can then be sought with high sensitivity.

The Hao et al. (2010) mixing diagram is a useful tool for selecting
non-standard quasar SEDs.

\section{Acknowledgments}
We thank Michael Strauss and Brandon Kelly for suggestions that
greatly improved the paper. This work was supported by NASA Chandra
grant number GO7-8136A (HH, ME, FC). This work is based in part on
data obtained as part of the UKIRT Infrared Deep Sky Survey.

\end{document}